\documentclass[10pt,journal,compsoc]{IEEEtran}
\ifCLASSOPTIONcompsoc
  \usepackage[nocompress]{cite}
\else
  \usepackage{cite}
\fi
\ifCLASSINFOpdf
\else
\fi

\makeatletter
\def\ps@headings{%
\def\@oddhead{\mbox{}\scriptsize\rightmark \hfil \thepage}%
\def\@evenhead{\scriptsize\thepage \hfil \leftmark\mbox{}}%
\def\@oddfoot{}%
\def\@evenfoot{}}
\makeatother
\usepackage{subfigure}
\usepackage{graphicx, amssymb, amsbsy, amsmath, amsthm}
\usepackage{multirow}
\usepackage{threeparttable}
\usepackage{cite,url}

\usepackage{booktabs}
\usepackage{color,soul}
\usepackage[font=footnotesize,labelfont=sf,textfont=sf]{caption}
\usepackage[lined,ruled,commentsnumbered]{algorithm2e}
\usepackage{algorithmic}
\usepackage{bbding}
\usepackage[colorlinks,linkcolor=red,anchorcolor=blue,citecolor=blue]{hyperref}
\graphicspath{{./figures/}}
\DeclareGraphicsExtensions{.pdf}
\UseRawInputEncoding

\def\squareforqed{\hbox{\rlap{$\sqcap$}$\sqcup$}}
\def\qed{\ifmmode\squareforqed\else{\unskip\nobreak\hfil
\penalty50\hskip1em\null\nobreak\hfil\squareforqed
\parfillskip=0pt\finalhyphendemerits=0\endgraf}\fi}

\begin{document}
\title{\emph{iGniter:} Interference-Aware GPU Resource Provisioning for Predictable DNN Inference in the Cloud}

\author{Fei~Xu,~\IEEEmembership{Member,~IEEE},
		Jianian~Xu,
		Jiabin Chen,
		Li~Chen,~\IEEEmembership{Member,~IEEE},
		Ruitao Shang,
        Zhi~Zhou,~\IEEEmembership{Member,~IEEE},
        Fangming~Liu,~\IEEEmembership{Senior Member,~IEEE}% <-this % stops a space
\IEEEcompsocitemizethanks{\IEEEcompsocthanksitem Fei Xu, Jianian Xu, Jiabin Chen, Ruitao Shang are with the Shanghai Key Laboratory of Multidimensional Information Processing, School of Computer Science and Technology, East China Normal University, 3663 N. Zhongshan Road, Shanghai 200062, China. Email: {\tt fxu@cs.ecnu.edu.cn}.
\IEEEcompsocthanksitem Li Chen is with the School of Computing and Informatics, University of Louisiana at Lafayette, 301 East Lewis Street, Lafayette, LA 70504, USA. E-mail: {\tt li.chen@louisiana.edu}.
\IEEEcompsocthanksitem Zhi Zhou is with the Guangdong Key Laboratory of Big Data Analysis and Processing, School of Computer Science and Engineering, Sun Yat-sen University, 132 E. Waihuan Road, Guangzhou 510006, China. E-mail: {\tt zhouzhi9@mail.sysu.edu.cn}.
\IEEEcompsocthanksitem Fangming Liu is with the National Engineering Research Center for Big Data Technology and System, the Services Computing Technology and System Lab, Cluster and Grid Computing Lab, School of Computer Science and Technology, Huazhong University of Science and Technology, 1037 Luoyu Road, Wuhan 430074, China. E-mail: {\tt fmliu@hust.edu.cn}.
%\IEEEcompsocthanksitem XX is with XX University.
}% <-this % stops an unwanted space
\thanks{Manuscript received January XX, 2022; revised April XX, 2022.}}


\markboth{Submitted to IEEE TRANSACTIONS ON PARALLEL AND DISTRIBUTED SYSTEMS}%
{Xu \MakeLowercase{\textit{et al.}}: \emph{iGniter:} Interference-Aware GPU Resource Provisioning for Predictable DNN Inference in the Cloud}
\IEEEtitleabstractindextext{%
\begin{abstract}

GPUs are essential to accelerating the latency-sensitive deep neural network (DNN) inference workloads in cloud datacenters. To fully utilize GPU resources, \emph{spatial sharing} of GPUs among co-located DNN inference workloads becomes increasingly compelling. However, GPU sharing inevitably brings \emph{severe performance interference} among co-located inference workloads, as motivated by an empirical measurement study of DNN inference on EC2 GPU instances. While existing works on guaranteeing inference performance service level objectives (SLOs) focus on either \emph{temporal sharing} of GPUs or \emph{reactive} GPU resource scaling and inference migration techniques, how to \emph{proactively} mitigate such severe performance interference has received comparatively little attention. In this paper, we propose \emph{iGniter}, an \emph{interference-aware} GPU resource provisioning framework for cost-efficiently achieving predictable DNN inference in the cloud. \emph{iGniter} is comprised of two key components: (1) a \emph{lightweight} DNN inference performance model, which leverages the system and workload metrics that are practically accessible to capture the performance interference; (2) A \emph{cost-efficient} GPU resource provisioning strategy that \emph{jointly} optimizes the GPU resource allocation and adaptive batching based on our inference performance model, with the aim of achieving predictable performance of DNN inference workloads. We implement a prototype of \emph{iGniter} based on the NVIDIA Triton inference server hosted on EC2 GPU instances. Extensive prototype experiments on four representative DNN models and datasets demonstrate that \emph{iGniter} can guarantee the performance SLOs of DNN inference workloads with practically acceptable runtime overhead, while saving the monetary cost by up to $25\%$ in comparison to the state-of-the-art GPU resource provisioning strategies.

\end{abstract}

}

\maketitle

\IEEEdisplaynontitleabstractindextext

% \fi
\IEEEpeerreviewmaketitle

\IEEEraisesectionheading{\section{Introduction}\label{sec:intro}}

\IEEEPARstart{W}{ith} the proliferating artificial intelligence applications, deep neural network (DNN) inference workloads are becoming increasingly commonplace in cloud datacenters~\cite{VYTJ2017}. While DNN models are getting more complex and thus consuming more computation and memory resources, GPUs have served as the \emph{key} accelerator to reduce the inference latency and meet the service level objective (SLO)~\cite{PXAHRAJI2018}. Hence, modern internet companies like Google, Alibaba, and JD are increasingly adopting GPUs for serving DNN inference in their latency-critical products such as voice assistants~\cite{intel-nvidia}, recommendation systems~\cite{GNYQWCXK2019}, and video analysis~\cite{jd}. To cut down the inference budget and facilitate cloud-based DNN inference, most cloud providers have recently launched commercial cloud AI platforms such as AWS SageMaker~\cite{EZBLBRNJAYP2020} and Google Vertex AI~\cite{vertexai}. As reported by Omdia, NVIDIA GPUs held an $80.6\%$ market share of AI processors in cloud datacenters in 2020 and expect to reach $37.6$ billion in revenue worldwide by 2026~\cite{omdia}.

To improve the utilization of GPU resources, \emph{temporal sharing}~\cite{HLYLBMAR2019} and \emph{spatial sharing}~\cite{YRR2021} are two common GPU resource multiplexing techniques. Many existing works (\emph{e.g.,} Cocktail~\cite{JCPMC2022}, Clockwork~\cite{ARSWAYJ2020}) leverage temporal sharing of GPUs to optimize the DNN inference performance and reduce the monetary cost. However, a recent study~\cite{ASK2020} has shown that temporal sharing of GPUs to execute DNN inference workloads can intrinsically result in GPU resource wastage. To \emph{fully exploit} the computation and memory resources of GPUs, NVIDIA has recently developed the multi-process service (MPS)~\cite{mps} technique, which allows multiple inference workloads to \emph{spatially} share the GPU resources with a limited percentage~\cite{WQNWKM2021} (\emph{e.g.,} $50\%$).

Though MPS can configure an amount of GPU resources for each inference workload, there exists \emph{noticeable performance interference} among the DNN inference workloads \emph{co-located} on a GPU device. As evidenced by our motivation experiments in Sec.~\ref{sec:motivation-interference}, the DNN inference latency can be prolonged by around $35\%$ with only $5$ co-located workloads on a GPU device. Such severe performance interference makes inference workloads easily suffer from unexpected SLO violations, which mainly originate from the shared \emph{resource contention} in three aspects: (1) the \emph{increased scheduling delay of kernels} by the GPU scheduler, and (2) the \emph{severe contention of GPU L2 cache space}, as well as (3) the \emph{reduced GPU frequency due to limited power cap}. Accordingly, it is essential to explicitly consider performance interference when provisioning GPU resources to DNN inference workloads, in order to meet the stringent performance SLOs for users.

\begin{figure}[!t]
	\centering\includegraphics[width=3.5in]{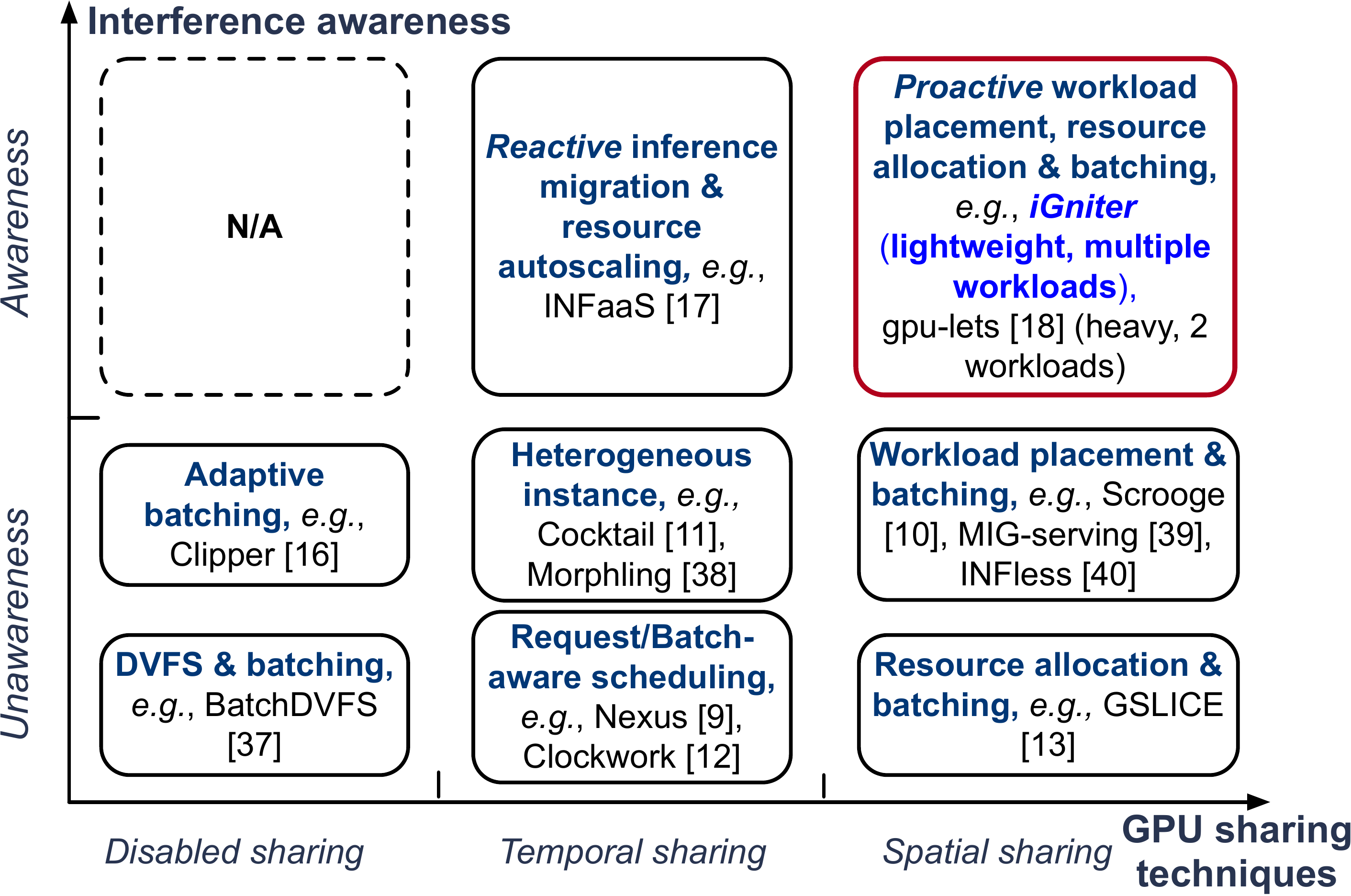}
	\caption{\emph{iGniter} positioning in the literature context of predictable DNN inference serving on GPUs.}
	\label{fig-igniter-positioning}\vspace{-10pt}
\end{figure}

To guarantee the performance SLOs of DNN inference workloads, many research efforts have been devoted to batch size configuration (\emph{e.g.,} Clipper~\cite{DXGMJI2017}), request scheduling (\emph{e.g.,} Clockwork~\cite{ARSWAYJ2020}), resource autoscaling (\emph{e.g.,}  Cocktail~\cite{JCPMC2022}), and GPU resource allocation (\emph{e.g.,}  GSLICE~\cite{ASK2020}), as summarized in Fig.~\ref{fig-igniter-positioning}. However, they are \emph{oblivious} to the severe performance interference among inference workloads, which is likely to cause resource under-provisioning and thus trigger \emph{frequent reactive} adjustment of GPU resources. There have also been recent works on mitigating such performance interference through \emph{reactive} inference migration (\emph{e.g.,} INFaaS~\cite{FQNC2021}) or characterizing the performance interference of \emph{two} co-located workloads using a linear regression model (\emph{e.g.,} gpu-lets~\cite{SSYJYJ2022}). Nevertheless, such an interference model requires a large number (\emph{i.e.,} thousands) of workload profiling and cannot readily be applied to multiple co-located inference workloads. As a result, there has been scant research attention paid to achieving predictable DNN inference by characterizing the performance interference in a \emph{lightweight} manner and \emph{proactively} mitigating such interference for inference workloads.

To fill this gap, in this paper, we design and implement \emph{iGniter}, an \emph{interference-aware} GPU resource provisioning framework to achieve predictable performance~\cite{FFHA2014} (\emph{i.e.,} latency and throughput) of DNN inference workloads while minimizing the inference budget in the cloud. To the best of our knowledge, \emph{iGniter} is the first attempt to demonstrate how to \emph{characterize the performance interference of DNN inference on GPUs in a lightweight manner, and cost-efficiently provision GPU resources for inference workloads by jointly optimizing the GPU resource allocation and adaptive batching.} Specifically, we make the following contributions in \emph{iGniter} as below.

$\vartriangleright$ First, \textbf{we build a \emph{lightweight} analytical performance model to explicitly capture the performance interference among DNN inference workloads} (Sec.~\ref{sec:model}). It empirically leverages a set of key system and workload metrics (\emph{e.g.,} the GPU L2 cache utilization, the number of kernels) to characterize the severe contention of GPU scheduler, GPU L2 cache space, and GPU power consumption, as identified by our motivation experiments in Sec.~\ref{sec:motivation-interference}.

$\vartriangleright$ Second, \textbf{we propose a \emph{cost-efficient} GPU resource provisioning strategy to guarantee the performance SLOs of DNN inference workloads} (Sec.~\ref{sec:design-algorithm}). Given the DNN models with their performance SLOs, \emph{iGniter} first leverage our inference performance model to calculate the appropriate batch size and lower bound of allocated GPU resources. It then greedily identifies the GPU device for placement with the minimum performance interference and allocates GPU resources for each inference workload.

$\vartriangleright$ Finally, \textbf{we implement a prototype\footnote{\href{https://github.com/icloud-ecnu/igniter}{https://github.com/icloud-ecnu/igniter}} of \emph{iGniter} based on the NVIDIA Triton inference server}~\cite{triton} with three pieces of modules, including an \emph{inference workload placer} and a \emph{GPU resource allocator} as well as an \emph{inference performance predictor} (Sec.~\ref{sec:design-implement}). We conduct prototype experiments on a cluster of $10$ p3.2xlarge GPU instances with $12$ representative inference workloads on Amazon EC2 (Sec.~\ref{sec:evaluation}). Experiment results show that \emph{iGniter} delivers predictable performance to DNN inference workloads with acceptable runtime overhead, while reducing the monetary cost by up to $25\%$ compared with the state-of-the-art GPU resource provisioning strategies.

%The rest of the paper is organized as follows. Sec.~\ref{sec:motivation} empirically analyzes the key factors that cause performance interference of inference workloads co-located on GPUs with MPS enabled, which motivates the design of our analytical performance model of DNN inference workloads in Sec.~\ref{sec:model}. Sec.~\ref{sec:design} further designs and implements our \emph{iGniter} GPU resource provisioning strategy for achieving predictable DNN inference serving in the cloud. Sec.~\ref{sec:evaluation} evaluates the effectiveness and runtime overhead of \emph{iGniter}. Sec.~\ref{sec:related} discusses related work and Sec.~\ref{sec:conclusion} concludes this paper.

\section{Background and Motivation}
\label{sec:motivation}

In this section, we first seek to analyze the severity of performance interference among co-located DNN inference workloads and identify the key factors that cause such interference. Next, we present an illustrative example to show how to adequately provision GPU resources for workloads to achieve predictable DNN inference.

\subsection{Multi-Process Service of NVIDIA GPUs}
\label{sec:motivation-mps}

To provide powerful computing ability, the NVIDIA GPU has been equipped with a number of Streaming Multiprocessors (SMs), and accordingly, GPUs are currently widely used for hosting DNN inference workloads in the cloud~\cite{ARSWAYJ2020}. To improve the resource utilization of GPUs, NVIDIA MPS~\cite{mps} has been developed to share GPU resources (\emph{i.e.,} SMs) among multiple inference workloads executed on a single GPU device. One process commonly hosts one inference workload. However, an uncontrollable allocation of GPU resources can degrade the Quality-of-Service (QoS) of DNN inference workloads. To deal with such a performance issue, MPS provisions each DNN inference workload with an amount of limited GPU resources (\emph{i.e.,} a set of SMs), starting from the NVIDIA Volta architecture~\cite{WQNWKM2021}. In general, the batch size of DNN inference also requires tuning to improve the GPU resource utilization, without violating the performance SLOs of inference workloads~\cite{PXAHRAJI2018}.

\begin{figure*}
	\begin{minipage}[t]{0.32\linewidth}
		\centering
		\includegraphics[width=2.27in]{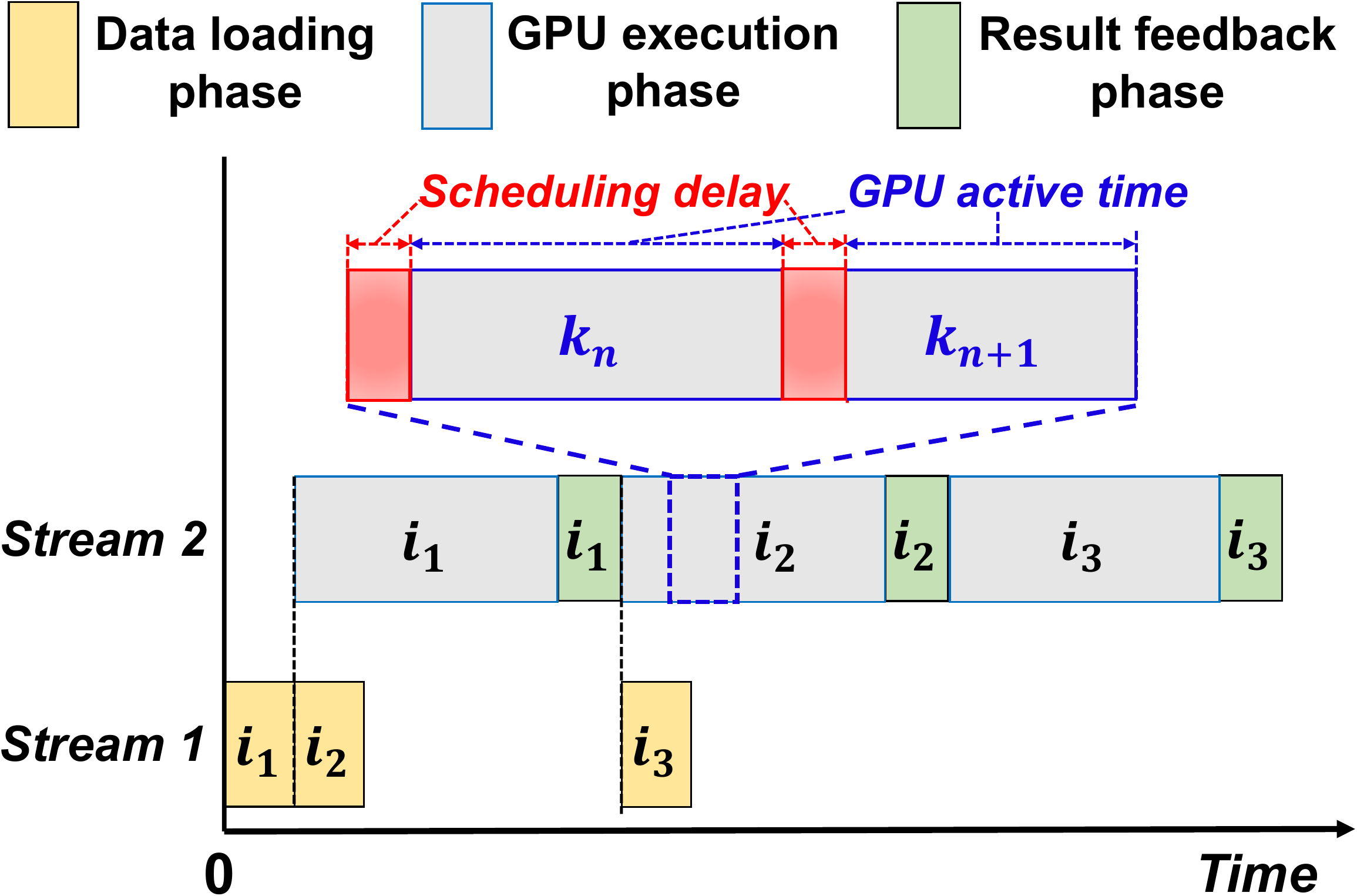}
		\caption{CUDA streams mechanism overlaps the execution of different DNN inference queries (\emph{i.e.,} $i_{1}, i_{2}, i_{3}$) in an inference workload, and the kernels (\emph{e.g.,} $k_{n}$) are scheduled onto SMs during the GPU execution phase.}
		\label{fig-motivation-inferenceprocess}
	\end{minipage}\hspace{+6pt}
	\begin{minipage}[t]{0.32\linewidth}
		\centering
		\includegraphics[width=2.0in]{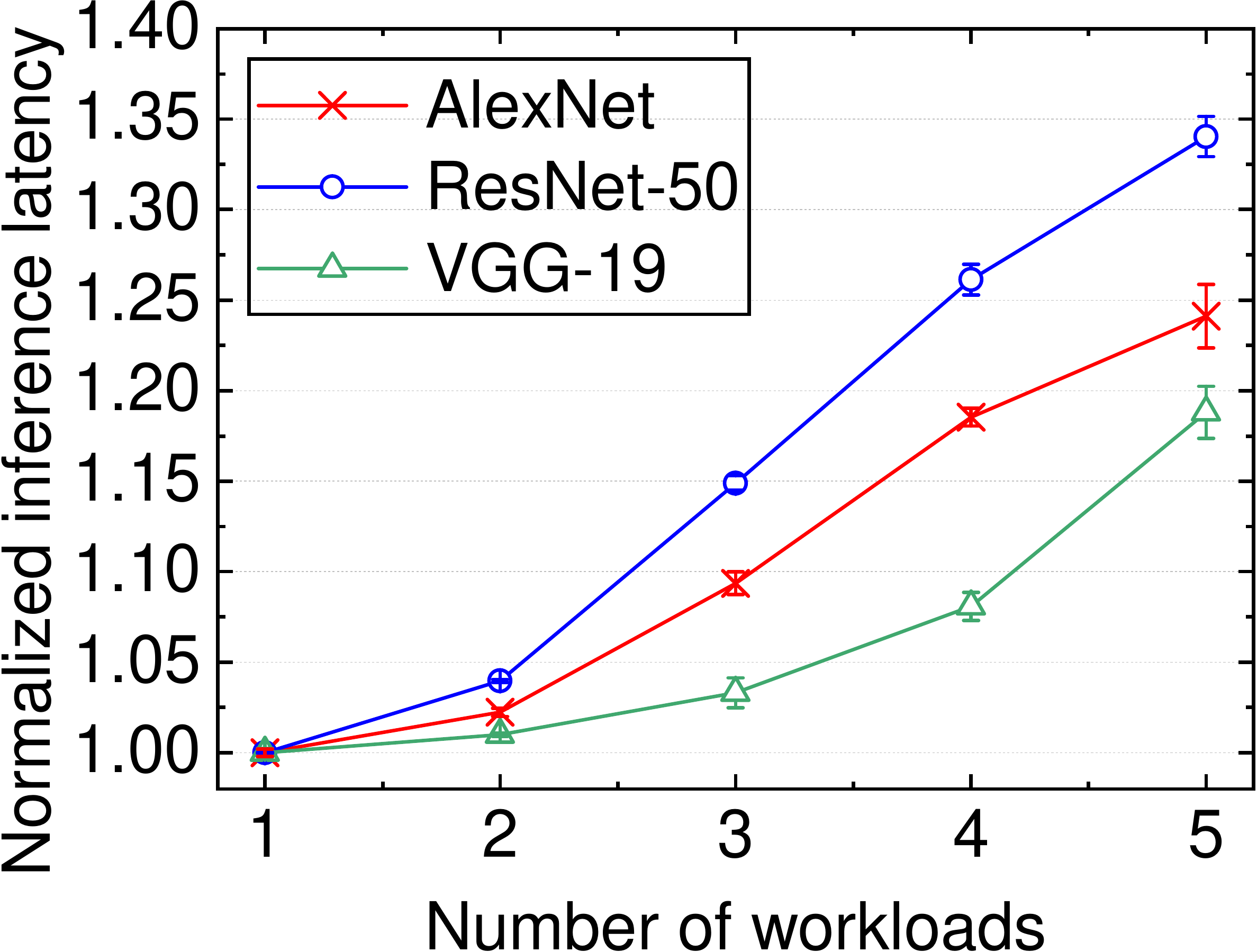}
		\caption{Normalized inference latency of AlexNet, ResNet-50, and VGG-19 achieved on a V100 GPU, as the number of co-located inference workloads varies from $1$ to $5$, with respect to the workloads running alone.}
		\label{fig-motivation-processesinterference}
	\end{minipage}\hspace{+6pt}
	\begin{minipage}[t]{0.32\linewidth}
		\centering
		\includegraphics[width=2.0in]{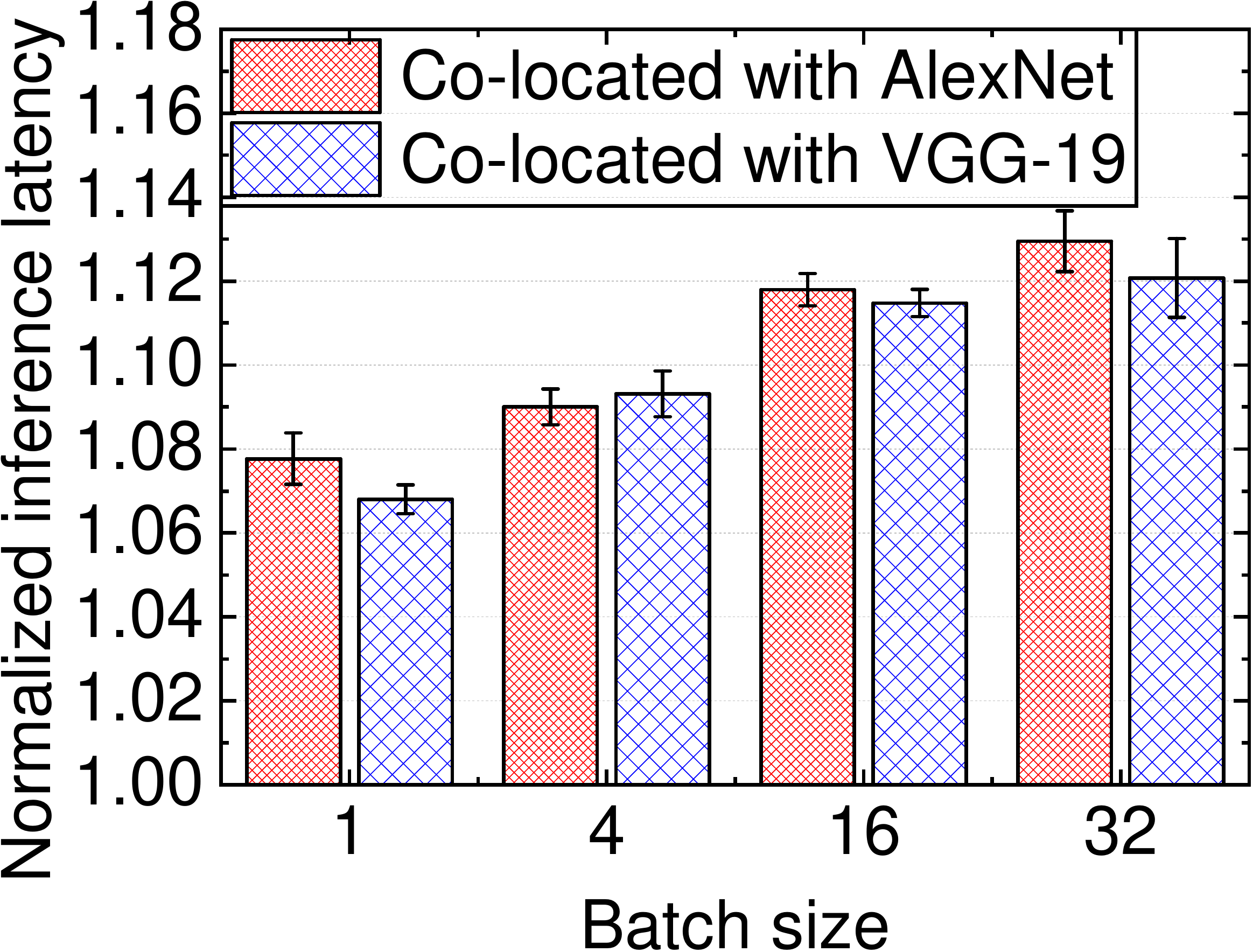}
		\caption{Normalized inference latency of ResNet-50 when co-located with AlexNet or VGG-19 on a V100 GPU, as the batch sizes of AlexNet and VGG-19 vary from $1$ to $32$, with respect to ResNet-50 running alone.}
		\label{fig-motivation-batchinterference}	
	\end{minipage}\vspace{-10pt}
\end{figure*}

The execution of a DNN inference workload on a GPU device mainly has three phases: \emph{First,} the host CPU transmits the inference input data to the GPU device over the PCIe interconnect. \emph{Second,} the GPU device executes the DNN inference query. \emph{Finally,} the inference result is transmitted back to the host CPU via the PCIe interconnect. To improve the GPU resource utilization, the mainstream DNN inference servers (\emph{e.g.,} NVIDIA Triton~\cite{triton}) have developed the CUDA \emph{streams} to \emph{overlap} the data loading phase and the GPU execution phase of different DNN inference queries in an \emph{asynchronous} manner. As shown in Fig.~\ref{fig-motivation-inferenceprocess}, the DNN inference queries (\emph{i.e.,} $i_{1}, i_{2}, i_{3}$) are launched in two different streams which can be executed concurrently. Specifically, Stream 1 (\emph{i.e.,} the data loading phase of $i_{2}$ and $i_{3}$) overlaps with Stream 2 (\emph{i.e.,} the GPU execution phase of $i_{1}$ and $i_{2}$). In particular, an inference query consists of a number of kernels (\emph{e.g.,} $k_{n}$) which require \emph{scheduling} onto SMs~\cite{SSY2021}, leading to a moderate amount of \emph{scheduling delay} of kernels in the GPU execution stream.

\subsection{Performance Interference among Co-located DNN Inference Workloads}
\label{sec:motivation-interference}

Though MPS facilitates the spatial GPU resource sharing among co-located inference workloads, it still brings \emph{non-negligible} performance interference. To examine the \emph{severity} of such interference, we conduct two motivation experiments using p3.2xlarge EC2 instances~\cite{ec2} equipped with NVIDIA V100 GPUs. We use AlexNet~\cite{AIG2017}, ResNet-50~\cite{KXSJ2016}, and VGG-19~\cite{KA2015} models executed on the NVIDIA TensorRT~\cite{H2016} framework as our DNN inference workloads. Specifically, we \emph{first} launch $1$ to $5$ identical inference workloads concurrently and each is allocated $20\%$ of GPU resources. \emph{Second,} we launch two DNN inference workloads on a GPU, and each is allocated $50\%$ of GPU resources. We vary the batch size of one workload from $1$ to $32$ while fixing the batch size of the other workload as $16$. In particular, we measure the \emph{average} DNN inference latency by excluding the inference batching delay. We illustrate the experimental results with error bars of standard deviation by repeating each experiment three times.

As shown in Fig.~\ref{fig-motivation-processesinterference} and Fig.~\ref{fig-motivation-batchinterference}, the DNN inference latency increases from $0.83\%$ to $34.98\%$, as the number of co-located workloads increases from $2$ to $5$ and the batch size of co-located inference workloads varies from $1$ to $32$. The experiment results indicate that the performance interference is \emph{not uncommon} for MPS even with limited GPU resources (\emph{i.e.,} GPU spatial sharing~\cite{mps}). Our observation above is consistent with the findings in a more recent work~\cite{SSYJYJ2022}. Through an in-depth analysis, we find that such severe performance interference among DNN inference workloads is mainly caused by the following three factors.

\begin{figure*}
	\begin{minipage}[t]{0.32\linewidth}
		\centering
		\includegraphics[width=2.0in]{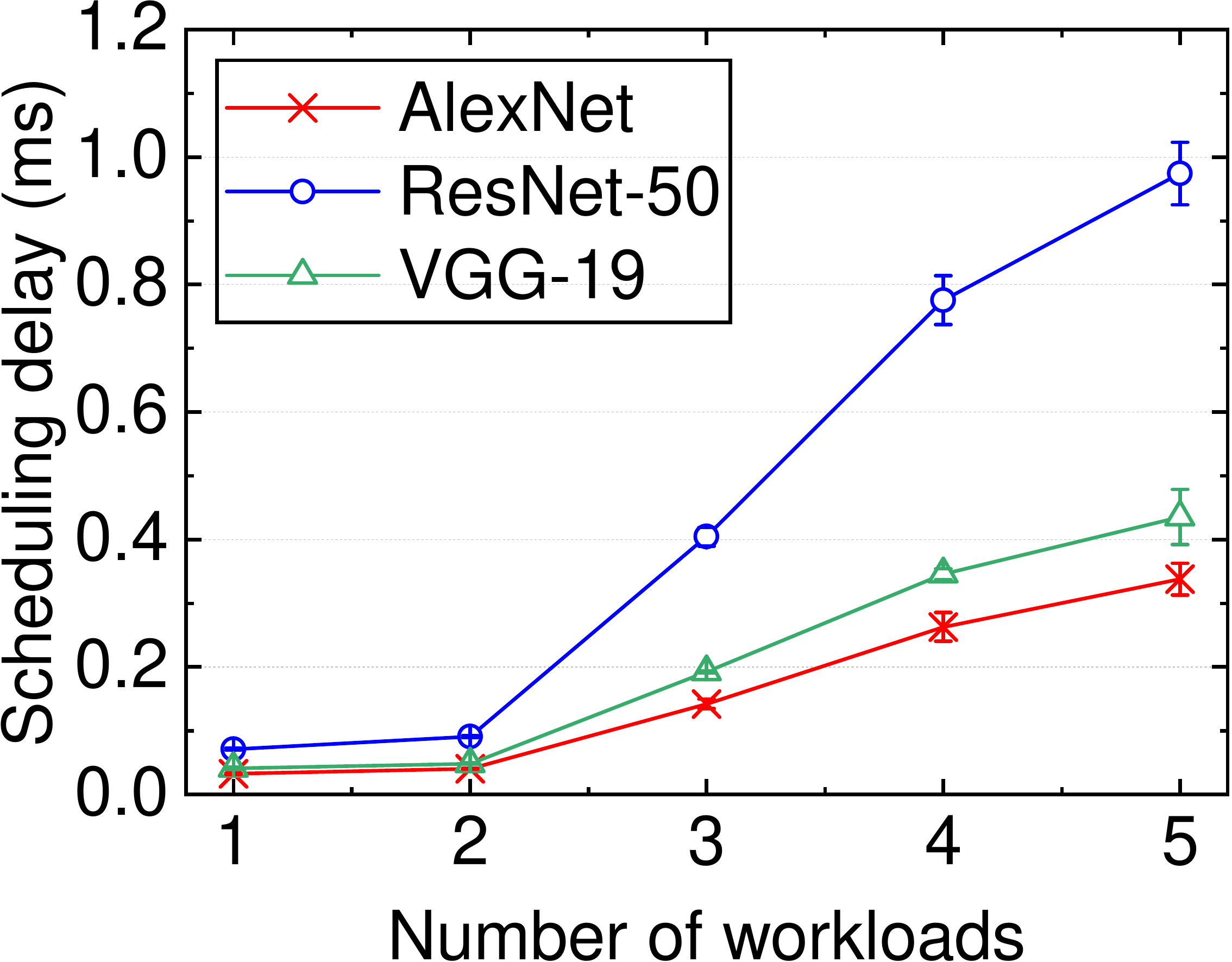}
		\caption{Scheduling delay of AlexNet, ResNet-50, and VGG-19 with different numbers of workloads executed on a V100 GPU.}
		\label{fig-motivation-idletime}
	\end{minipage}\hspace{+3pt}
\begin{minipage}[t]{0.32\linewidth}
	\centering
		\includegraphics[width=2.36in]{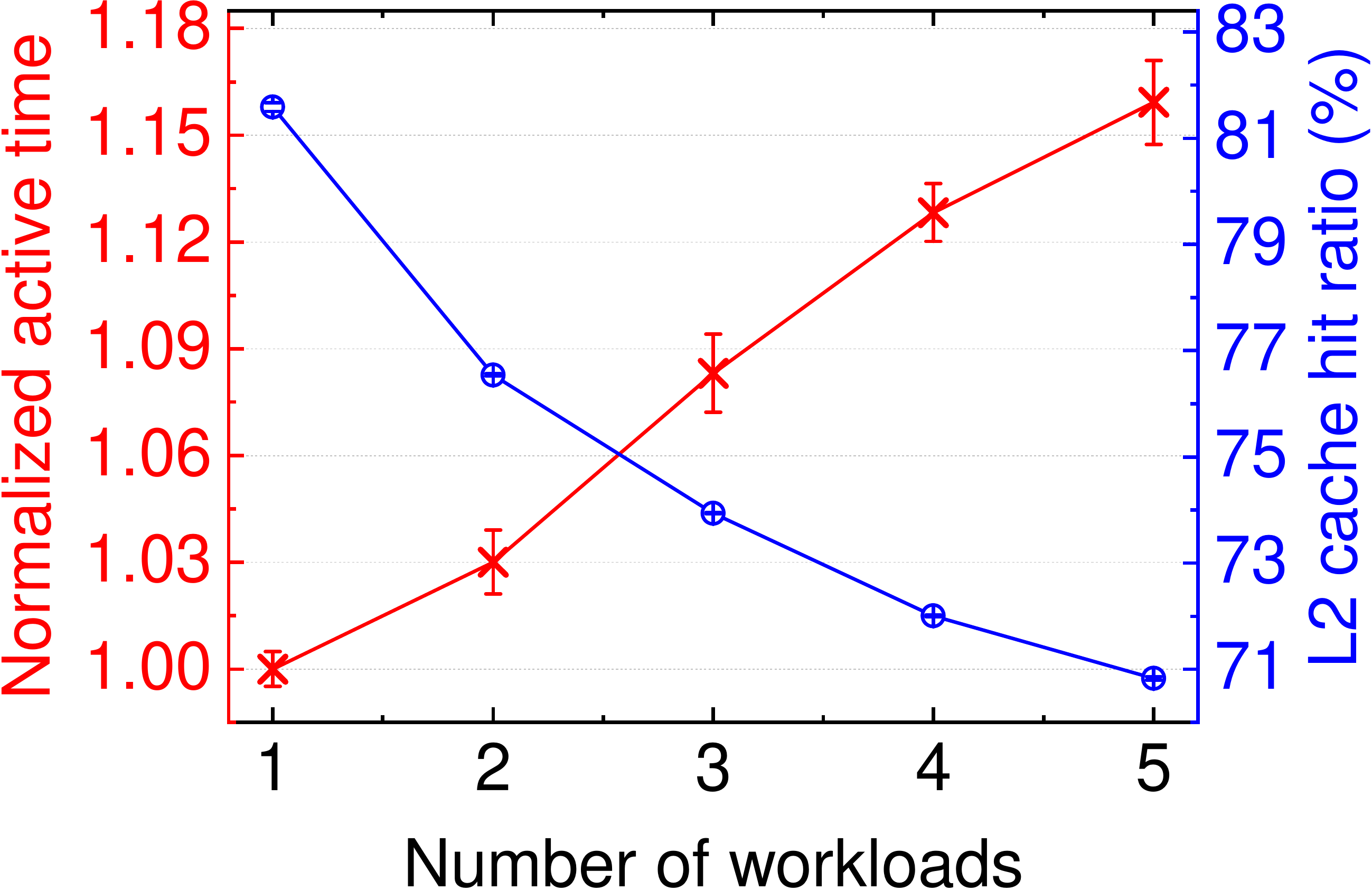}
		\caption{GPU active time and L2 cache request hit ratio of ResNet-50 with different numbers of workloads executed on a V100 GPU.}
		\label{fig-motivation-l2cachehitrate}
	\end{minipage}\hspace{+10pt}
	\begin{minipage}[t]{0.32\linewidth}
		\centering
		\includegraphics[width=2.42in]{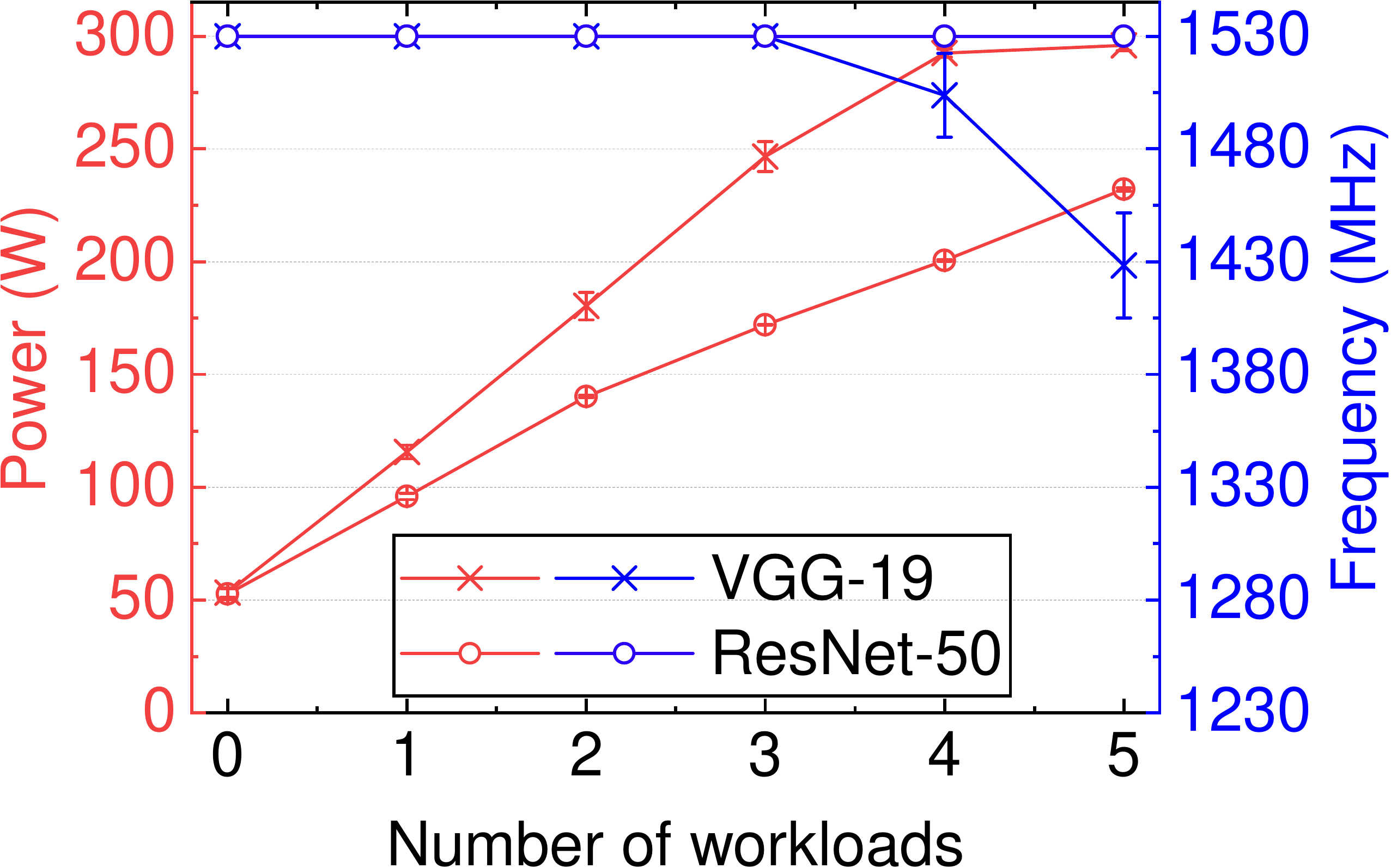}
		\caption{GPU power and GPU frequency of VGG-19 and ResNet-50 with different numbers of workloads executed on a V100 GPU.}
		\label{fig-motivation-power}
	\end{minipage}\vspace{-10pt}
\end{figure*}

\emph{Increased Scheduling Delay of Kernels.} Each kernel of a DNN inference workload needs to be scheduled onto SMs by the GPU scheduler. As shown in Fig.~\ref{fig-motivation-idletime}, we observe that: \emph{First,} the scheduling delay shows a roughly linear increase as the number of co-located workloads increases from $2$ to $5$. We conjecture that the GPU scheduler requires scheduling the kernels from different inference workloads onto SMs in a round-robin manner. \emph{Second,} the scheduling delay of ResNet-50 increases much faster than AlexNet. This is simply because the number of kernels of ResNet-50 is bigger than that of AlexNet.

\emph{Severe Contention of GPU L2 Cache Space.} Though MPS can partition GPU resources, the GPU L2 cache space is still shared by co-located DNN inference workloads~\cite{SISR2019}. To characterize the severity of such L2 cache contention on a GPU device, we simply adopt a system metric, \emph{i.e.,} the L2 cache request hit ratio. As shown in Fig.~\ref{fig-motivation-l2cachehitrate}, we observe that the GPU active time (\emph{i.e.,} GPU execution latency - GPU scheduling delay, as depicted in Fig.~\ref{fig-motivation-inferenceprocess}) of ResNet-50 is inversely related to the GPU L2 cache hit ratio. As the number of co-located workloads increases, the severer cache contention leads to a smaller L2 cache hit ratio, which in turn increases the GPU active time of an inference workload.

\emph{Reduced GPU Frequency due to Limited Power Cap.} Reduction of GPU frequency brings performance degradation to GPU workloads~\cite{RRJAMZ2013}. As shown in Fig.~\ref{fig-motivation-power}, we observe that: \emph{First,} the GPU frequency starts to decrease once the GPU power reaches its upper limit value. This is because more inference workloads consume a larger amount of power on a GPU device, while the GPU has to maintain the upper limit of GPU power through frequency reduction. \emph{Second,} the GPU power of VGG-19 and ResNet-50 shows a roughly linear relationship to the number of inference workloads, as long as the GPU power is below its upper limit value.

Based on our analysis above, we further explain why the batch size of co-located workloads (\emph{i.e.,} AlexNet, VGG-19) can \emph{moderately affect} the DNN inference performance (\emph{i.e.,} ResNet-50) by $6.36\%$ -- $13.93\%$, as shown in Fig.~\ref{fig-motivation-batchinterference}. Such performance interference can mainly be attributed to the resource contention of GPU L2 cache space and GPU power. As the batch sizes of AlexNet and VGG-19 increase from $1$ to $32$, the GPU L2 cache utilization of the two workloads increases from $11.1\%$ to $18.4\%$ and from $16.9\%$ to $22.0\%$, respectively. Similarly, the GPU power of AlexNet and VGG-19 also increases from $108$ W to $156$ W and from $139$ W to $179$ W, respectively, thereby causing GPU frequency reduction. Accordingly, such severe contention of the GPU L2 cache space and GPU power from co-located inference workloads inevitably prolongs the DNN inference latency.

\textbf{Summary.} \emph{First,} the performance interference among DNN inference workloads cannot be overlooked. We identify the main factors that cause such interference as the severe contention of the GPU scheduler, GPU L2 cache space, and GPU power consumption among co-located inference workloads on a GPU device. \emph{Second,} explicitly considering the performance interference is compelling when provisioning GPU resources to DNN inference workloads, so as to guarantee the performance of DNN inference workloads.

\subsection{An Illustrative Example}
\label{sec:motivation-example}

To achieve predictable DNN inference performance and cost-efficient GPU resource provisioning, we propose \emph{iGniter} in Sec.~\ref{sec:design} and illustrate its effectiveness by conducting another motivation experiment with AlexNet, ResNet-50, and VGG-19 models. We set the latency SLOs (ms) and request arrival rates (req/s) for the three inference workloads as $15$, $40$, $60$ and $500$, $400$, $200$, respectively. We define the P99 latency of an inference workload exceeding its latency SLO as a violation.

\begin{table}[!t]\vspace{+6pt}
\renewcommand{\arraystretch}{1.3}
\centering
\caption{Comparison of GPU resource provisioning plans and SLO violations achieved by the gpu-lets, GSLICE and our \emph{iGniter} strategies for three representative DNN models (\emph{i.e.,} AlexNet ($\mathtt{A}$), ResNet-50 ($\mathtt{R}$), VGG-19 ($\mathtt{V}$)).}
\label{table-motivation-example}
\begin{tabular}{ccc}
\toprule[1pt]
\multirow{2}{*}{Approaches} & {Resource provisioning plans} & \multirow{2}{*}{Violations}\\
\cline{2-2}
  & $\mathtt{GPU}$: $\mathtt{model (\text{\#}resource, \text{\#}batch)}$\\
\midrule[1pt]
\multirow{2}{*}{GSLICE~\cite{ASK2020}} & {$\mathtt{GPU1: A (37.5\%, 18)}$,} & $2$ models\\
& $\mathtt{R (30\%, 8), V (40\%, 6)}$ & ($\mathtt{A, R}$)\\
\hline
\multirow{2}{*}{gpu-lets~\cite{SSYJYJ2022}} & {$\mathtt{GPU1: A (40\%, 23)}$} & $2$ models \\
\cline{2-2}
 & $\mathtt{GPU2: R (60\%, 18), V (40\%, 6)}$ & ($\mathtt{A, R}$)\\
\hline
\multirow{2}{*}{\emph{iGniter}} & {$\mathtt{GPU1: A (10\%, 4)}$,} &  \multirow{2}{*}{None}\\
& $\mathtt{R (30\%, 8), V (37.5\%, 6)}$ \\
\bottomrule[1pt]
\end{tabular}\vspace{-10pt}
\end{table}

As shown in Table~\ref{table-motivation-example}, GSLICE~\cite{ASK2020} and gpu-lets~\cite{SSYJYJ2022} require $1$ GPU and $2$ GPUs, respectively. Unfortunately, they make two DNN models violate their SLOs. In contrast, our \emph{iGniter} strategy provisions $1$ GPU for hosting the three models appropriately and it guarantees their SLOs. Specifically, we find that GSLICE and gpu-lets tend to provision more GPU resources and larger batch sizes to AlexNet and ResNet-50 than \emph{iGniter}. This is because the two strategies aim to maximize the request throughput while guaranteeing latency SLOs. In addition, GSLICE~\cite{ASK2020} is an \emph{interference-unaware} strategy, which tunes the allocated GPU resources for inference workloads \emph{separately}. Accordingly, the total allocated resources can exceed the maximum resources (\emph{i.e.,} $100$\%) of a GPU device which inevitably leads to the contention of SMs, causing high long-tail inference latency.

Though gpu-lets~\cite{SSYJYJ2022} explicitly considers the performance interference, it works \emph{only for two inference workloads} on a GPU device. Also, gpu-lets only considers the interference for the \emph{newly-arrived} inference workload (\emph{i.e.,} VGG-19), and it does not change the allocated GPU resources and batch size of the \emph{originally-placed} workload (\emph{i.e.,} ResNet-50) on the GPU. Accordingly, the inference latency of ResNet-50 exceeds its latency SLO due to the interference impact from VGG-19. Moreover, gpu-lets first provisions an \emph{efficient} amount of GPU resources and then sets the batch size as large as possible for inference workloads. However, a large batch size cannot fully utilize the GPU resources at a low request arrival rate. It can cause SLO violations due to long batching latency. In contrast, \emph{iGniter} sets an appropriate batch size for inference workloads that \emph{just meet} their latency SLOs and request arrival rates. It further provisions GPU resources by explicitly considering the interference among multiple (more than $2$) inference workloads to guarantee the DNN inference performance in a cost-efficient manner.

\section{Modeling DNN Inference Performance on GPUs}
\label{sec:model}

In this section, we first build an analytical model to predict the DNN inference performance in the cloud. We explicitly consider the performance interference among DNN inference workloads with different batch sizes and allocated GPU resources. We next formulate the GPU resource provisioning problem to minimize the monetary cost while guaranteeing inference performance SLOs. The key notations in our performance model are summarized in Table~\ref{table-notations}.

\begin{table}[!t]\vspace{+0pt}
\renewcommand{\arraystretch}{1.3}\vspace{+5pt}
\centering \caption{Key notations in our DNN inference performance model.} \label{table-notations}\vspace{-0pt}
\begin{tabular}{c|p{7.1cm}}
\toprule[1pt]
Notation & Definition\\
\midrule[1pt]
$\mathcal{I}, \mathcal{J}$ & Sets of DNN inference workloads and allocated GPUs\\
\hline
\multirow{2}{*}{$t_{inf}^{ij}$} & DNN inference latency of an inference workload $i$ on a GPU $j$\\
\hline
$h^{ij}$ & Throughput of an inference workload $i$ on a GPU $j$\\
\hline
$t_{load}^{i}$, & DNN inference data loading latency and result\\
$t_{feedback}^{i}$ & feedback latency of an inference workload $i$\\
\hline
\multirow{2}{*}{$t_{gpu}^{ij}$} & GPU execution latency of an inference workload $i$ on a GPU $j$\\
\hline
\multirow{2}{*}{$t_{sch}^{ij}$, $t_{act}^{ij}$} & Scheduling delay and GPU active time of an inference workload $i$ on a GPU $j$\\
\hline
$f^{j}$ & Actual frequency of a GPU $j$\\
\hline
$p_{demand}^{j}$ & Total power demand of a GPU $j$\\
\hline
\multirow{2}{*}{$k_{act}^{i}$} & GPU active time of an inference workload $i$ when running alone on a GPU device\\
\hline
\multirow{2}{*}{$p^{i}$, $c^{i}$} & Power consumption and L2 cache utilization of an inference workload $i$ when running alone on a GPU device\\
\hline
\multirow{2}{*}{$r^{ij}$, $v^{ij}$} & GPU resource allocation and placement of an inference workload $i$ on a GPU $j$\\
\hline
$b^{i}$ & Batch size of an inference workload $i$\\
\bottomrule[1pt]
\end{tabular}
\vspace{-10pt}
\end{table}

\subsection{Predicting DNN Inference Performance with GPU Resources}
\label{sec:model-performance}

We consider a set of \emph{constantly-arrived} DNN inference workloads denoted by $\mathcal{I} = \{i_{1},i_{2},...,i_{m}\}$ over a period of time (\emph{e.g.,} several minutes). A set of GPU devices to be allocated is denoted by $\mathcal{J} = \{j_{1},j_{2},...,j_{g}\}$ with \emph{a given GPU type}. As elaborated in Sec.~\ref{sec:motivation-mps}, the execution of DNN inference on the GPU can be divided into three sequential steps: \emph{data loading}, \emph{GPU execution}, and \emph{result feedback}. Accordingly, the DNN inference latency $t_{inf}^{ij}$ of a workload $i$ executed on a GPU device $j$ can be calculated by summing up the data loading latency $t_{load}^{i}$, the GPU execution latency $t_{gpu}^{ij}$, and the result feedback latency $t_{feedback}^{i}$, which is given by
\begin{equation}\label{eq-inference-latency}
	t_{inf}^{ij} = t_{load}^{i} + t_{gpu}^{ij} + t_{feedback}^{i}.
\end{equation}
As discussed in Sec.~\ref{sec:motivation-mps}, the data loading phase \emph{overlaps} with the GPU execution and result feedback phases in the mainstream DNN inference servers (\emph{e.g.,} Triton~\cite{triton}) to improve the GPU resource utilization. Accordingly, we estimate the DNN inference throughput $h^{ij}$ as
\begin{equation}\label{eq-throughput}
	h^{ij} = \frac{b^{i}}{t_{gpu}^{ij} + t_{feedback}^{i}},
\end{equation}
where $b^{i} \in \mathcal{N}^{+}$ denotes the batch size of an inference workload $i \in \mathcal{I}$.

\emph{Data Loading and Result Feedback Phases.} As discussed in Sec.~\ref{sec:motivation-mps}, the inference input and result data are transmitted between the CPU and GPU devices via the PCIe. In general, both the inference input data size and result data are linear to the batch size $b^{i}$. We calculate the data loading latency $t_{load}^{i}$ and the result feedback latency $t_{feedback}^{i}$ as
\begin{equation}\label{eq-pcie-latency}
	t_{load}^{i} = \frac{d_{load}^{i} \cdot b^{i}}{B_{pcie}} \quad \text{and} \quad t_{feedback}^{i} = \frac{d_{feedback}^{i} \cdot b^{i}}{B_{pcie}},
\end{equation}
respectively, where $d_{load}^{i}$ and $d_{feedback}^{i}$ are the input data size and result data size, respectively, when $b^{i} = 1$. $B_{pcie}$ denotes the available PCIe bandwidth of a GPU device.

\emph{GPU Execution Phase.} Each DNN inference workload is executed with an amount of allocated GPU resources denoted by $r^{ij} \in [0, r_{max}], \forall i \in \mathcal{I}, j \in \mathcal{J}$, which are actually mapped to a set of SMs~\cite{mps}. In general, $r_{max}$ is set as $1$. As depicted in Fig.~\ref{fig-motivation-inferenceprocess}, the GPU execution phase consists of GPU scheduling and kernels running on the allocated SMs (\emph{i.e.,} $r^{ij}$). Moreover, the GPU execution phase can be prolonged by the GPU frequency reduction due to the workload co-location, as evidenced by Sec.~\ref{sec:motivation-interference}. Accordingly, we formulate the GPU execution latency $t_{gpu}^{ij}$ as
\begin{equation}\label{eq-gpu-latency}
	t_{gpu}^{ij} = \frac{t_{sch}^{ij} + t_{act}^{ij}}{\frac{f^{j}}{F}},
\end{equation}
where $t_{sch}^{ij}$ and $t_{act}^{ij}$ denote the total scheduling delay of kernels and the GPU active time of an inference workload $i$ executed on a GPU device $j$, respectively, \emph{without any GPU frequency reductions.} $f^{j}$ and $F$ denote the actual and maximum GPU frequency, respectively, on a GPU device $j$.

In the following, we first model the \emph{scheduling delay} $t_{sch}^{ij}$ of DNN inference workloads. Intuitively, $t_{sch}^{ij}$ is roughly linear to the number of kernels $n_{k}^{i}$ for a DNN inference workload $i$, which can be estimated as
\begin{equation}\label{eq-schedule-latency}
	t_{sch}^{ij} = \big(k_{sch}^{i} + \Delta_{sch}^{j}\big) \cdot n_{k}^{i},
\end{equation}
where $k_{sch}^{i}$ denotes the scheduling delay when the workload $i$ is running alone on a GPU device. $\Delta_{sch}^{j}$ is the increased scheduling delay caused by the interference on the GPU resource scheduler, which is relevant to the number of co-located inference workloads as evidenced by Sec.~\ref{sec:motivation-interference}. Accordingly, we estimate the increased scheduling delay as
\begin{equation}\label{eq-increased-schedule-latency}
	\Delta_{sch}^{j} = \left\{
		\begin{array}{lll}
			0  & & \sum\limits_{i \in \mathcal{I}}v^{ij} \leq 1, \\
			\alpha_{sch} \cdot \sum\limits_{i \in \mathcal{I}}v^{ij} + \beta_{sch} & & \text{otherwise},
		\end{array}
		\right.
\end{equation}
where $\alpha_{sch}$ and $\beta_{sch}$ are the coefficients to characterize the increased scheduling delay on a given GPU type. $\sum_{i \in \mathcal{I}}v^{ij}$ denotes the number of co-located inference workloads on a GPU device $j$. $v^{ij}$ denotes whether an inference workload $i$ is running on a GPU device $j$, which is given by
\begin{equation}\label{eq-whether-running}
	v^{ij} = \left\{
		\begin{array}{lll}
			1 \quad \text{a workload $i$ runs on a GPU $j$ ($r^{ij} > 0$)},\\
			0 \quad \text{otherwise ($r^{ij} = 0$)}.
		\end{array}
		\right.
\end{equation}

We next model the \emph{GPU active time} $t_{act}^{ij}$ of an inference workload $i$ executed on a GPU device $j$. As evidenced by Sec.~\ref{sec:motivation-interference}, the GPU active time is inversely proportional to the GPU L2 cache hit ratio. We simply leverage a system metric called \emph{GPU L2 cache utilization} to characterize the workload \emph{demand} on the GPU L2 cache space. Given a fixed \emph{supply} of L2 cache space on a GPU device, a higher GPU L2 cache utilization (\emph{i.e.,} \emph{demand}) indicates severer contention on the GPU L2 cache space, thereby causing a longer GPU active time. Accordingly, we estimate $t_{act}^{ij}$ as
\begin{equation}\label{eq-cache-interference}
	t_{act}^{ij} = k_{act}^{i} \cdot \Big(1 + \alpha_{cache}^{i} \cdot \sum_{i \in \mathcal{I} \setminus i}\big(c^{i} \cdot v^{ij}\big)\Big),
\end{equation}
where $\alpha_{cache}^{i}$ denotes the coefficient to characterize the prolonged GPU active time due to L2 cache contention for an inference workload $i$. $k_{act}^{i}$ and $c^{i}$ are the GPU active time and L2 cache utilization, respectively, when an inference workload $i$ is running alone on a GPU device.

Finally, we model the \emph{GPU frequency} $f^{j}$ on a GPU device $j$. As evidenced by Sec.~\ref{sec:motivation-interference}, the GPU frequency decreases dramatically as the total GPU power \emph{demand} $p_{demand}^{j}$ of workloads exceeds the upper limit of GPU power \emph{supply} $P$ of a GPU device. As the GPU frequency is highly relevant to the GPU power~\cite{RRJAMZ2013}, we estimate $f^{j}$ as
\begin{equation}\label{eq-frequency}
	f^{j} = \left\{
		\begin{array}{lll}
			F  & & p_{demand}^{j} \leq P,\\
			F + \alpha_{f} \cdot \big(p_{demand}^{j} - P\big)  & & p_{demand}^{j} > P,
		\end{array}
		\right.
\end{equation}
where $\alpha_{f}$ denotes the coefficient to characterize the relationship between the GPU power and frequency on a GPU device. In addition, we estimate the total power demand of a GPU device $j$ by summing up the power consumption $p^{i}$ of all workloads and the \emph{idle} power $p_{idle}$ of a GPU device, which is given by
\begin{equation}\label{eq-power-demand}
	p_{demand}^{j} = p_{idle} + \sum\limits_{i \in \mathcal{I}}\big(p^{i} \cdot v^{ij}\big).
\end{equation}
In particular, we obtain $p^{i}$ by running an inference workload $i$ alone on a GPU device of the given type.

\textbf{Obtaining Model Coefficients.} Based on the above, we have $8$ \emph{workload-specific} coefficients (\emph{i.e.,} $d_{load}^{i}$, $d_{feedback}^{i}$, $n_{k}^{i}$, $k_{sch}^{i}$, $k_{act}^{i}$, $p^{i}$, $c^{i}$, $\alpha_{cache}^{i}$) and $7$ \emph{hardware-specific} coefficients (\emph{i.e.,} $P$, $F$, $p_{idle}$, $B_{pcie}$, $\alpha_{f}$, $\alpha_{sch}$, $\beta_{sch}$) in our performance model. Specifically, four workload-specific coefficients (\emph{i.e.,} $d_{load}^{i}$, $d_{feedback}^{i}$, $n_{k}^{i}$, $k_{sch}^{i}$) are obtained by profiling the workload \emph{only once} using the $\mathtt{Nsight}$ $\mathtt{Systems}$~\cite{nsight}. The available PCIe bandwidth $B_{pcie}$ is measured by transferring data from the main memory to GPU memory. Given a GPU type, three hardware-specific coefficients (\emph{i.e.,} $P$, $F$, $p_{idle}$) are obtained using the $\mathtt{nvidia-smi}$~\cite{nvidia-smi}. The GPU frequency coefficient $\alpha_{f}$ and scheduling coefficients ($\alpha_{sch}$, $\beta_{sch}$) as well as cache coefficient $\alpha_{cache}^{i}$ are obtained by launching multiple (\emph{e.g.,} $2$ to $5$) inference workloads concurrently. Moreover, we obtain the GPU active time $k_{act}^{i}$, power consumption $p^{i}$, and the L2 cache utilization $c^{i}$ of an inference workload $i$ \emph{running alone on a GPU device} as follows.

\begin{figure}[!t]
    \begin{minipage}[t]{0.47\linewidth}
		\centering
		\includegraphics[width=1.8in]{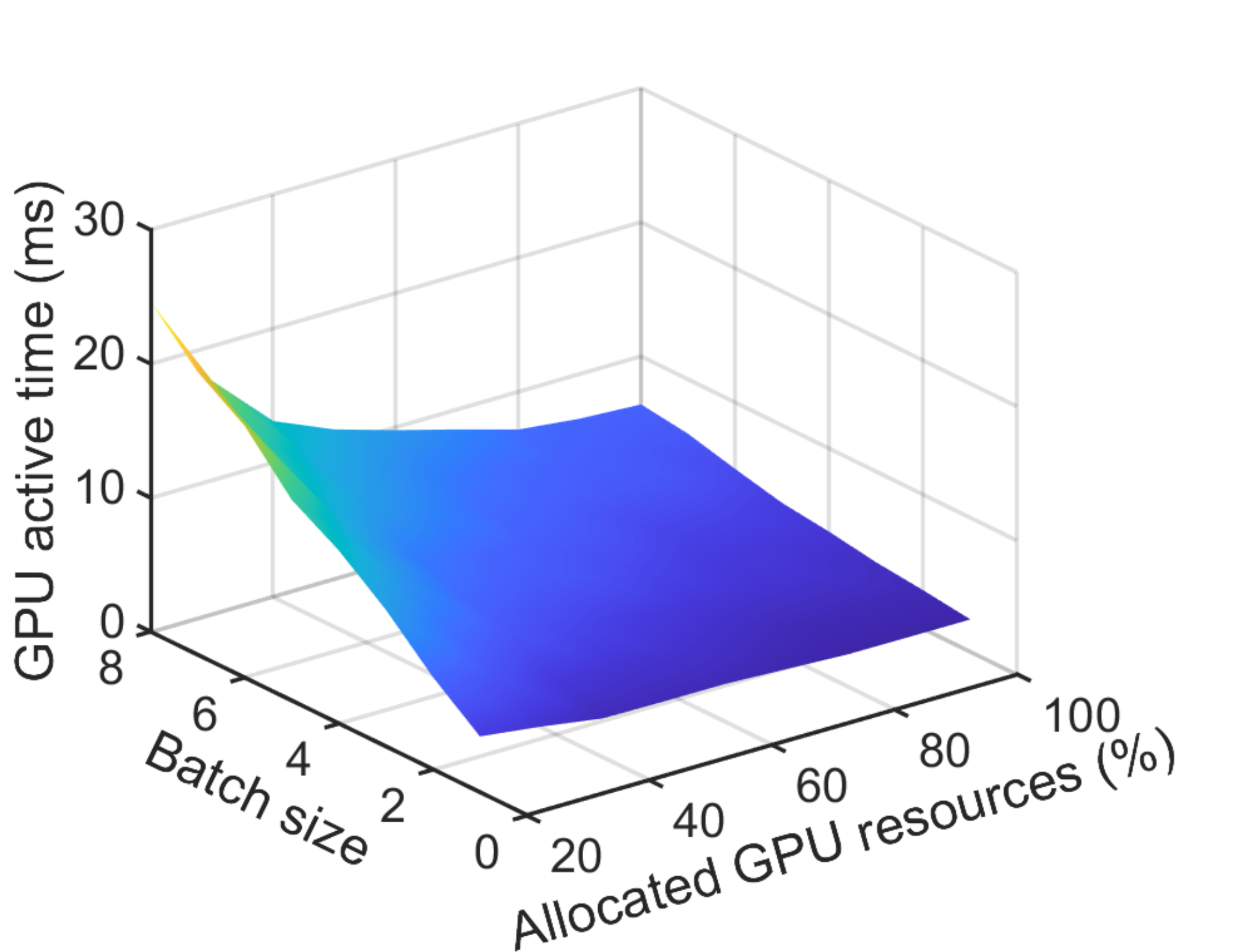}
		\caption{GPU active time of ResNet-50 with different batch sizes and allocated GPU resources.}
	\label{fig-model-activetime}
	\end{minipage}
	\hspace{+3pt}
	\begin{minipage}[t]{0.47\linewidth}
		\centering
		\includegraphics[width=1.8in]{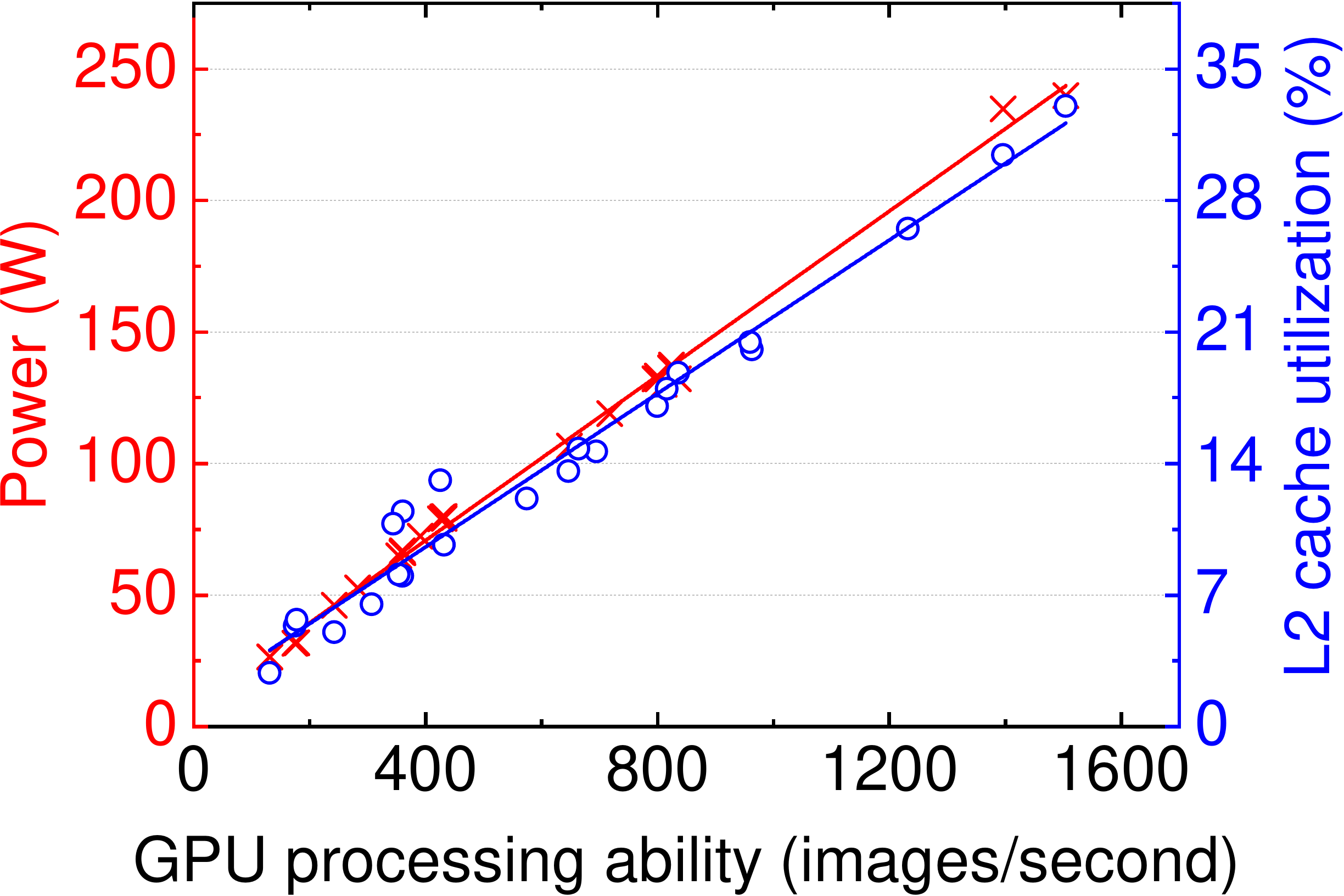}
		\caption{Power consumption and L2 cache utilization of ResNet-50 with different GPU processing abilities.}
		\label{fig-model-power-l2cache}
	\end{minipage}\vspace{-15pt}
\end{figure}

Specifically, as depicted in Fig.~\ref{fig-model-activetime}, the GPU active time $k_{act}^{i}$ shows a roughly inverse proportion to the amount of allocated GPU resources $r^{ij}$. Also, the GPU active time increases fast with the batch size $b^{i}$, which can be formulated by a quadratic function. Accordingly, we formulate $k_{act}^{i}$ as
\begin{equation}\label{eq-single-gpu-latency}
	k_{act}^{i} = \frac{k_{1}^{i} \cdot (b^{i})^{2} + k_{2}^{i} \cdot b^{i} + k_{3}^{i}}{r^{ij}+k_{4}^{i}}+k_{5}^{i},
\end{equation}
where $k_{1}^{i}$, $k_{2}^{i}$, $k_{3}^{i}$, $k_{4}^{i}$, $k_{5}^{i}$ denote the model coefficients for an inference workload $i$. In addition, Fig.~\ref{fig-model-power-l2cache} shows that both the power consumption $p^{i}$ and L2 cache utilization $c^{i}$ (measured by $\mathtt{Nsight}$ $\mathtt{Compute}$~\cite{nsight-compute}) of an inference workload $i$ grow \emph{linearly} with the GPU processing ability (\emph{i.e.,} $\frac{b}{k_{act}}$). This is because a stronger GPU processing ability commonly leads to higher GPU resource utilization and power consumption. Accordingly, we estimate $p^{i}$ and $c^{i}$ as
\begin{eqnarray}
	p^{i} &=& \alpha_{power}^{i} \cdot \frac{b^{i}}{k_{act}^{i}} + \beta_{power}^{i}, \label{eq-solo-power}\nonumber\\
	c^{i} &=& \alpha_{cacheutil}^{i} \cdot \frac{b^{i}}{k_{act}^{i}} + \beta_{cacheutil}^{i}, \label{eq-solo-cache}\nonumber
\end{eqnarray}
where $\alpha_{power}^{i}$, $\beta_{power}^{i}$ and $\alpha_{cacheutil}^{i}$, $\beta_{cacheutil}^{i}$ denote the model coefficients to characterize the relationship between the power consumption, L2 cache utilization and the GPU processing ability. Such model coefficients above can be obtained by fitting several (\emph{e.g.,} more than $5$) sets of profiled workload data using the \emph{least squares method}~\cite{H2007}. In particular, we only require profiling each inference workload with $11$ different configurations of allocated GPU resources and batch sizes, which is far less than the number (\emph{i.e., $40 \times 32 = 1,280$}) of all possible configurations of allocated GPU resources (\emph{e.g.,} $40$ choices) and batch sizes (\emph{e.g.,} $32$ choices) for each inference workload, even without considering performance interference.

\subsection{Analyzing GPU Resource Provisioning Optimization Problem}
\label{sec:model-problem}

Based on our DNN inference performance model above, we proceed to define the optimization problem of GPU resource provisioning as follows: \emph{Given the inference performance SLOs in terms of the request arrival rate $R^{i}$ and latency SLO $T_{slo}^{i}$, how can we provision GPU resources $r^{ij}$ and configure batch size $b^{i}$ for each inference workload $i$, to achieve predictable DNN inference performance while minimizing the monetary cost $C$ of allocated GPU resources?} Accordingly, our online optimization problem can be formulated as
\begin{eqnarray}
	\min_{b^{i}, r^{ij}} & & C = \sum_{j \in \mathcal{J}}u^{j} \label{eq-optimization} \\
	\text{s.t.} & & \sum_{j \in \mathcal{J}}h^{ij} \cdot v^{ij} \geq R^{i}, \quad \forall i \in \mathcal{I} \label{eq-cons-arrivalrate}\\
            & & \sum_{j \in \mathcal{J}}t_{inf}^{ij} \cdot v^{ij} \leq \frac{T_{slo}^{i}}{2}, \quad \forall i \in \mathcal{I} \label{eq-cons-slo}\\
            & & \sum_{i \in \mathcal{I}}r^{ij} \leq r_{max}, \quad \forall j \in \mathcal{J} \label{eq-cons-resource}\\
            & & \sum_{j \in \mathcal{J}}v^{ij} = 1, \quad \forall i \in \mathcal{I} \label{eq-cons-inference-number}
            %& &	v \in \mathcal{V}, m \in \mathcal{M}, b \in \mathcal{B}, sm \in \mathcal{SM}_{v}, i_{v} \in \mathcal{I} \label{eq-cons-number}
\end{eqnarray}
where $u^{j}$ denotes the unit price of each GPU device $j$, and Eq.~(\ref{eq-optimization}) defines our objective function which minimizes the monetary cost $C$ of GPU resource provisioning, subject to the following four constraints. Specifically, Constraint~(\ref{eq-cons-arrivalrate}) guarantees that the throughput of each inference workload can meet its arrival rate $R^{i}$. Constraint~(\ref{eq-cons-slo}) guarantees the inference latency of each inference workload below its objective latency $\frac{T_{slo}^{i}}{2}$. This is because the batch inference latency cannot exceed half of the SLO~\cite{HLYLBMAR2019} by excluding the performance impact of request batching and queueing. Constraint~(\ref{eq-cons-resource}) denotes that the allocated GPU resources of each GPU device should be no more than the maximum GPU resources $r_{max}$. Constraint~(\ref{eq-cons-inference-number}) denotes that each inference workload can only be placed on one GPU device.

\textbf{Problem Analysis.} According to Eq.~(\ref{eq-optimization}), the monetary cost $C$ is affected by the unit price $u^{j}$ and set of allocated GPU devices $\mathcal{J}$, as the DNN inference models and requests \emph{arrive constantly}. As $u^{j}$ becomes a constant value $u$ given a GPU type, the optimization problem can be reduced to minimizing the number $|\mathcal{J}|$ of provisioned GPU devices. To achieve such a goal, each inference workload requires to be allocated GPU resources that \emph{just meet} the request arrival rate and latency SLOs.

\newtheorem{theorem}{Theorem}
\begin{theorem}\label{thm-lower-upper-bound}
Given a DNN inference workload with the arrival rate and latency SLO, the lower bound $r_{lower}^{i}$ of allocated GPU resources (i.e., the allocated GPU resources that DNN inference workloads are running alone on a GPU device) and the appropriate batch size $b_{appr}^{i}$ can be calculated as
\begin{eqnarray}
    b_{appr}^{i} &=& \bigg\lceil \frac{T_{slo}^{i} \cdot R^{i} \cdot B_{pcie}}{2 \cdot \big(B_{pcie} + R^{i} \cdot  d_{load}^{i}\big)} \bigg\rceil,\label{eq-appropriate-batch-size}\\
    r_{lower}^{i} &=& \bigg\lceil \frac{\gamma^{i}}{\delta^{i} \cdot r_{unit}} - \frac{k_{4}^{i}}{r_{unit}} \bigg\rceil \cdot r_{unit}.\label{eq-lowerbound-resources}
\end{eqnarray}
where $\gamma^{i} = k_{1}^{i} \cdot (b_{appr}^{i})^2 + k_{2}^{i} \cdot b_{appr}^{i} +k_{3}^{i}$ and $\delta^{i} = \frac{T_{slo}^{i}}{2} -\frac{(d_{load}^{i} + d_{feedback}^{i}) \cdot b_{appr}^{i}}{B_{pcie}}-k_{5}^{i}-k_{sch}^{i}\cdot n_{k}^{i}$. $r_{unit}$ denotes the allocation unit of GPU resources, which can be empirically set as $2.5\%$ (i.e., around $2$ SMs) for NVIDIA V100 GPUs.
\end{theorem}

The proof can be found in Appendix~\ref{sec:appendix}. Our selected \emph{appropriate} batch size $b_{appr}^{i}$ can guarantee the request arrival rate by letting $t_{gpu}^{ij} = \frac{T_{slo}^{i}}{2} - t_{load}^{i} - t_{feedback}^{i}$. Accordingly, Constraint~(\ref{eq-cons-arrivalrate}) and Constraint~(\ref{eq-cons-slo}) can be combined as one constraint. The original optimization problem in Eq.~(\ref{eq-optimization}) can be simplified as
\begin{eqnarray}
	\min_{r^{ij}} & & \frac{u}{r_{max}} \cdot \Big(\sum_{i \in \mathcal{I}} r_{lower}^{i} + \sum_{j \in \mathcal{J}} \sum_{i \in \mathcal{I}}r_{inter}^{ij} + \sum_{j \in \mathcal{J}}r_{f}^{j}\Big) \\
	\text{s.t.}
            & & \frac{\big(d_{load}^{i} + d_{feedback}^{i}\big) \cdot b_{appr}^{i}}{B_{pcie}}+\sum_{j \in \mathcal{J}}t_{gpu}^{ij} \leq \frac{T_{slo}^{i}}{2}, \; \forall i \in \mathcal{I} \nonumber \\
            & &(\ref{eq-cons-resource}), \; (\ref{eq-cons-inference-number}), \nonumber
\end{eqnarray}
where $r_{inter}^{ij} = r^{ij} - r_{lower}^{i} \cdot v^{ij}$ is the \emph{increased} GPU resources caused by the interference of co-located inference workloads. $r_{f}^{j} = r_{max} - \sum_{i \in \mathcal{I}}r^{ij}$ denotes the \emph{unallocated} GPU resource fragments on a GPU device $j$. Accordingly, given the fixed lower bound $r_{lower}^{i}$ of GPU resources, our optimization problem can be transformed into minimizing the \emph{GPU resource fragmentation} and the \emph{increased GPU resources} caused by the performance interference. Suppose that there is no performance interference among the inference workloads (\emph{i.e.,} $r_{inter}^{ij} = 0$), our problem can be reduced to a classic \emph{bin packing problem} which is already shown to be NP-hard~\cite{J1973}. Obviously, our original optimization problem is more \emph{complicated} than such a bin packing problem. Accordingly, we turn to devising a heuristic algorithm to acquire an appropriate (\emph{i.e., sub-optimal}) solution to our GPU resource provisioning problem.

\section{Design of \emph{iGniter}: Guaranteeing Performance of DNN Inference Workloads}
\label{sec:design}

Based on the analysis of our DNN inference performance model and the optimization problem defined in Sec.~\ref{sec:model}, we further present \emph{iGniter} in Alg.~\ref{alg-inference-provisioning}, a \emph{simple yet effective} GPU resource provisioning strategy to provide predictable performance (\emph{i.e.,} guarantee the latency SLO and request arrival rate) for inference workloads, while minimizing the monetary cost of provisioned GPU resources in the cloud.

\subsection{Algorithm Design}
\label{sec:design-algorithm}

To particularly answer ``\emph{how to provision GPU resources for a set of DNN inference workloads},'' our \emph{iGniter} strategy in Alg.~\ref{alg-inference-provisioning} is quite \emph{intuitive}: We first decide \emph{where to place} inference workloads and then identify \emph{how to allocate} GPU resources to the workloads. To particularly reduce the unallocated GPU resource fragments, \emph{iGniter} sorts the inference workloads according to $r_{lower}^{i}$ in \emph{descending} order. It puts these workloads onto a new GPU device \emph{only when} there are not enough GPU resources, accordingly to the $\mathtt{ANYFIT}$ constraint~\cite{J1973}.

\SetAlFnt{\small}
\SetAlgoVlined \vspace{-0pt}
\renewcommand{\algorithmicrequire}{\textbf{Input:}}
\renewcommand{\algorithmicensure}{\textbf{Output:}}
\begin{algorithm}[!t]
\caption{\emph{iGniter:} Cost-efficient GPU resource provisioning strategy for achieving predictable performance of DNN inference workloads.}
\label{alg-inference-provisioning}
\begin{algorithmic}[1]
\REQUIRE The latency SLO $T_{slo}^{i}$ and the request arrival rate $R^{i}$ of each inference workload $i \in \mathcal{I}$.
\ENSURE Cost-efficient resource provisioning plan, including the provisioned GPU resources $r^{ij}$ and the appropriate batch size $b_{appr}^{i}$ as well as the number of allocated GPUs $g$.
\STATE Acquire \emph{hardware-specific} coefficients $P$, $F$, $p_{idle}$, $B_{pcie}$, $\alpha_{f}$, $\alpha_{sch}$, $\beta_{sch}$ for a given GPU type, and obtain \emph{workload-specific} coefficients $d_{load}^{i}$, $d_{feedback}^{i}$, $n_{k}^{i}$, $k_{sch}^{i}$, $k_{act}^{i}$, $p^{i}$, $c^{i}$, $\alpha_{cache}^{i}$ through profiling each workload $i \in \mathcal{I}$;
\STATE \textbf{Initialize:} the appropriate batch size $b_{appr}^{i} \gets$ Eq.~(\ref{eq-appropriate-batch-size}), the lower bound of GPU resources $r_{lower}^{i} \gets$ Eq.~(\ref{eq-lowerbound-resources}), and $r^{ij} \gets 0$, $\forall i \in \mathcal{I}, \forall j \in \mathcal{J}$, as well as $g \gets 1$;
\STATE Sort workloads according to $r_{lower}^{i}$ in descending order;
\FORALL {workload $w$ in $\mathcal{I}$ to be placed on GPUs}
    \STATE \textbf{Initialize:} the allocated GPU resources $r_{a}^{ij} \gets r^{ij}$, $\forall i \in \mathcal{I}, \forall j \in \mathcal{J}$, after placing an inference workload $w$, and the minimum increased GPU resources caused by the \emph{performance interference} $r_{inter}^{min} \gets r_{max}$, for placing the workload $w$ on the GPU $q \gets -1$;
    \FORALL {GPU device $j$ in $[1,g]$}
        \STATE $r_{a}^{ij} \gets $ $\mathtt{alloc\_gpus}(T_{slo}^{i}, r_{a}^{ij}, r_{lower}^{w})$;
        \STATE Calculate the increased GPU resources caused by the \emph{performance interference} $r_{inter}^{ij} \gets r_{a}^{ij} - r^{ij}, \forall i \in \mathcal{I}$ on the GPU $j$;
        \IF {\big($\sum\limits_{i \in \mathcal{I}}r_{a}^{ij} \leq r_{max}$\big) \&\& \big($\sum\limits_{i \in \mathcal{I}}r_{inter}^{ij} < r_{inter}^{min}$\big)}
            \STATE Set $q \gets j$, and $r_{inter}^{min} \gets \sum\limits_{i \in \mathcal{I}}r_{inter}^{ij}$;
        \ENDIF
    \ENDFOR \tcp*{$\mathtt{find \; an \; appropriate \; GPU \; for \; a \; workload \; w}$}
    \IF {$q == -1$}
        \STATE Update $g \gets g+1$, and $r^{wg} \gets r_{lower}^{w}$ \tcp*{$\mathtt{add \; one \; GPU}$}
    \ELSE
        \STATE Update $r^{iq} \gets r_{a}^{iq}$, $\forall i \in \mathcal{I}$ \tcp*{$\mathtt{enough \; GPU \; resources}$}
    \ENDIF
\ENDFOR
%\RETURN the resource provisioning plan $r^{ij}, \forall i \in \mathcal{I}, \forall j \in \mathcal{J}$.
\end{algorithmic}
\end{algorithm}\vspace{-0pt}

\textbf{Inference Workload Placement Strategy.} Given a set of DNN inference workloads with their latency SLOs $T_{slo}^{i}$ and request arrival rates $R^{i}$, \emph{iGniter} first obtains the \emph{hardware-specific} coefficients (\emph{i.e.,} $P$, $F$, $p_{idle}$, $B_{pcie}$, $\alpha_{f}$, $\alpha_{sch}$, $\beta_{sch}$) and the \emph{workload-specific} coefficients (\emph{i.e.,} $d_{load}^{i}$, $d_{feedback}^{i}$, $n_{k}^{i}$, $k_{sch}^{i}$, $k_{act}^{i}$, $p^{i}$, $c^{i}$, $\alpha_{cache}^{i}$) for each inference workload using a \emph{lightweight} coefficient acquisition method elaborated in Sec.~\ref{sec:model-performance} (line $1$). With such obtained coefficients, \emph{iGniter} calculates the appropriate batch size $b_{appr}^{i}$ by Eq.~(\ref{eq-appropriate-batch-size}) and the lower bound of allocated GPU resources $r_{lower}^{i}$ by Eq.~(\ref{eq-lowerbound-resources}) (line $2$). By iterating over the \emph{sorted} inference workloads set $\mathcal{I}$, \emph{iGniter} greedily finds an \emph{appropriate} GPU device to host each workload (lines $3$-$12$). In more detail, \emph{iGniter} initializes the allocated GPU resources $r_{a}^{ij}$ after placing the inference workload on the GPU (lines $5$). For each candidate GPU, \emph{iGniter} first calculates the allocated GPU resources $r_{a}^{ij}$ and the increased resources $r_{inter}^{ij}$ by Alg.~\ref{alg-resource-allocation} (lines $6$-$8$). It then greedily identifies the \emph{appropriate} GPU $q$ which can host the inference workload and cause the \emph{least} performance interference $r_{inter}^{min}$ (lines $9$-$12$). Finally, \emph{iGniter} provisions a new GPU device if there are not enough resources for the inference workload $w$ (\emph{i.e.,} $q == -1$). Otherwise, it directly places such a workload $w$ onto the GPU device $q$ with the minimum increased GPU resources (lines $13$-$18$).

\textbf{GPU Resource Allocation Strategy.} $\mathtt{alloc\_gpus}$ first initializes the allocated GPU resources $r_{a}^{wj}$ of the workload $w$ as $r_{lower}^{w}$ on the GPU $j$ (line $1$). $\mathtt{alloc\_gpus}$ then iteratively \emph{reallocates} the GPU resources for each workload $i$ on the GPU $j$, as long as SLO violations still occur for an inference workload $i$ and the GPU $j$ has enough unallocated GPU resources (lines $2$-$11$). Specifically, $\mathtt{alloc\_gpus}$ calculates the inference latency $t_{inf}^{ij}$ by Eq.~(\ref{eq-inference-latency}) and judges whether the SLO violation occurs for each workload $i$ (lines $4$-$6$). For these SLO-violated workloads, $\mathtt{alloc\_gpus}$ increases the allocated GPU resources by a unit of GPU resources (\emph{i.e.,} $r_{unit}$) to guarantee the inference SLOs (lines $7$-$11$).

\SetAlFnt{\small}
\SetAlgoVlined \vspace{-0pt}
\renewcommand{\algorithmicrequire}{\textbf{Input:}}
\renewcommand{\algorithmicensure}{\textbf{Output:}}
\begin{algorithm}[!t]
\caption{$\mathtt{alloc\_gpus}$: GPU resource allocation algorithm for placing an inference workload on a GPU device.}
\label{alg-resource-allocation}
\begin{algorithmic}[1]
\REQUIRE The latency SLO $T_{slo}^{i}$ and the allocated GPU resources $r_{a}^{ij}$ of each inference workload $i \in \mathcal{I}$, before placing the inference workload $w$ on the GPU $j$, as well as the resource lower bound $r_{lower}^{w}$ of the inference workload $w$.
\ENSURE Allocated GPU resources $r_{a}^{ij}$, after placing the inference workload ${w}$ on the GPU $j$.
\STATE \textbf{Initialize:} the allocated GPU resources $r_{a}^{wj} \gets r_{lower}^{w}$ of the workload $w$ on the GPU $j$, and whether the GPU resources require reallocation $flag \gets 1$;
\WHILE {\big($\sum\limits_{i \in \mathcal{I}}r_{a}^{ij} \leq r_{max}$\big) \&\& \big($flag == 1$\big)}
    \STATE \textbf{Initialize:} $flag \gets 0$;
    \FORALL {inference workload $i$ on the GPU $j$}
    	\STATE Calculate the inference latency $t_{inf}^{ij} \gets$ Eq.~(\ref{eq-inference-latency});
        \IF {$t_{inf}^{ij} > \frac{T_{slo}^{i}}{2}$}
            \STATE Increase the allocated GPU resources $r_{a}^{ij} \gets r_{a}^{ij} + r_{unit}$ for a workload $i$;
            \STATE Set $flag \gets 1$;
        \ENDIF \tcp*{$\mathtt{SLO \; violation \; occurs}$}
    \ENDFOR \tcp*{$\mathtt{Reallocate \; GPU \; resources}$}
\ENDWHILE
%\RETURN allocated GPU resources $r_{a}^{ij}, \forall i \in \mathcal{I}$ on the GPU $j$.
\end{algorithmic}
\end{algorithm}\vspace{-0pt}

\textbf{Remark.} As Alg.~\ref{alg-inference-provisioning} (line $7$) invokes Alg.~\ref{alg-resource-allocation}, the time and space complexities of Alg.~\ref{alg-inference-provisioning} are in the order of $\mathcal{O}(m \cdot g \cdot n \cdot \frac{m}{g})$ and $\mathcal{O}(m)$, respectively, where $m$ denotes the number of inference workloads and $g$ denotes the number of allocated GPUs. Also, $n = \frac{r_{max} - \sum_{i \in \mathcal{I}}r_{a}^{ij}}{r_{unit}} + 1$ denotes the cardinality of searching space of the allocated GPU resources for an inference workload. $\frac{m}{g}$ denotes the \emph{expected} number of inference workloads co-located on a GPU. As $n$ is practically limited (\emph{i.e.,} at most $40$ values in the real-world scenario), the time complexity of Alg.~\ref{alg-inference-provisioning} can be reduced to $\mathcal{O}(m^{2})$. To reduce the memory consumption of \emph{iGniter}, we store the sparse matrix $r^{ij}$ in Alg.~\ref{alg-inference-provisioning} and Alg.~\ref{alg-resource-allocation} using \emph{adjacency lists}, and accordingly the space complexities of Alg.~\ref{alg-inference-provisioning} can be in the order of $\mathcal{O}(m)$. As a result, the runtime and memory overhead of our \emph{iGniter} strategy is well contained and will be validated in Sec.~\ref{sec:eval-overhead}.

In particular, \emph{iGniter} can be generalized to the heterogeneous types of cloud instances (with different types of GPU hardware). Given multiple types of GPU instances, we only need to obtain the \emph{hardware-specific} coefficients and a part of \emph{workload-specific} coefficients (\emph{i.e.,} $k_{sch}^{i}$, $k_{act}^{i}$, $p^{i}$, $c^{i}$, $\alpha_{cache}^{i}$ in line 1 of Alg.~1) for each type of GPU device. The rest of Alg.~1 can directly be executed without any modifications. Accordingly, \emph{iGniter} can be easily extended to the heterogeneous cluster, by judiciously selecting the \emph{most cost-efficient type of GPU instances} for DNN inference workloads, which will be validated in Sec.~\ref{sec:eval-effectiveness}.

\subsection{Implementation of \emph{iGniter}}
\label{sec:design-implement}

We implement a prototype of the \emph{iGniter} framework running on Amazon EC2 GPU instances~\cite{ec2} based on NVIDIA Triton~\cite{triton}, which is a representative cloud inference server. More specifically, our \emph{iGniter} prototype is built upon the Triton server v2.12.0 supported by the TensorRT backend framework v8.0.1.6, with over $1,000$ lines of Python, C++, and Linux Shell codes. The source codes of our \emph{iGniter} prototype are publicly available on GitHub (\emph{i.e.,} \href{https://github.com/icloud-ecnu/igniter}{$\mathtt{https://github.com/icloud-ecnu/igniter}$}).

\begin{figure}[!t]
	\centering\includegraphics[width=3.4in]{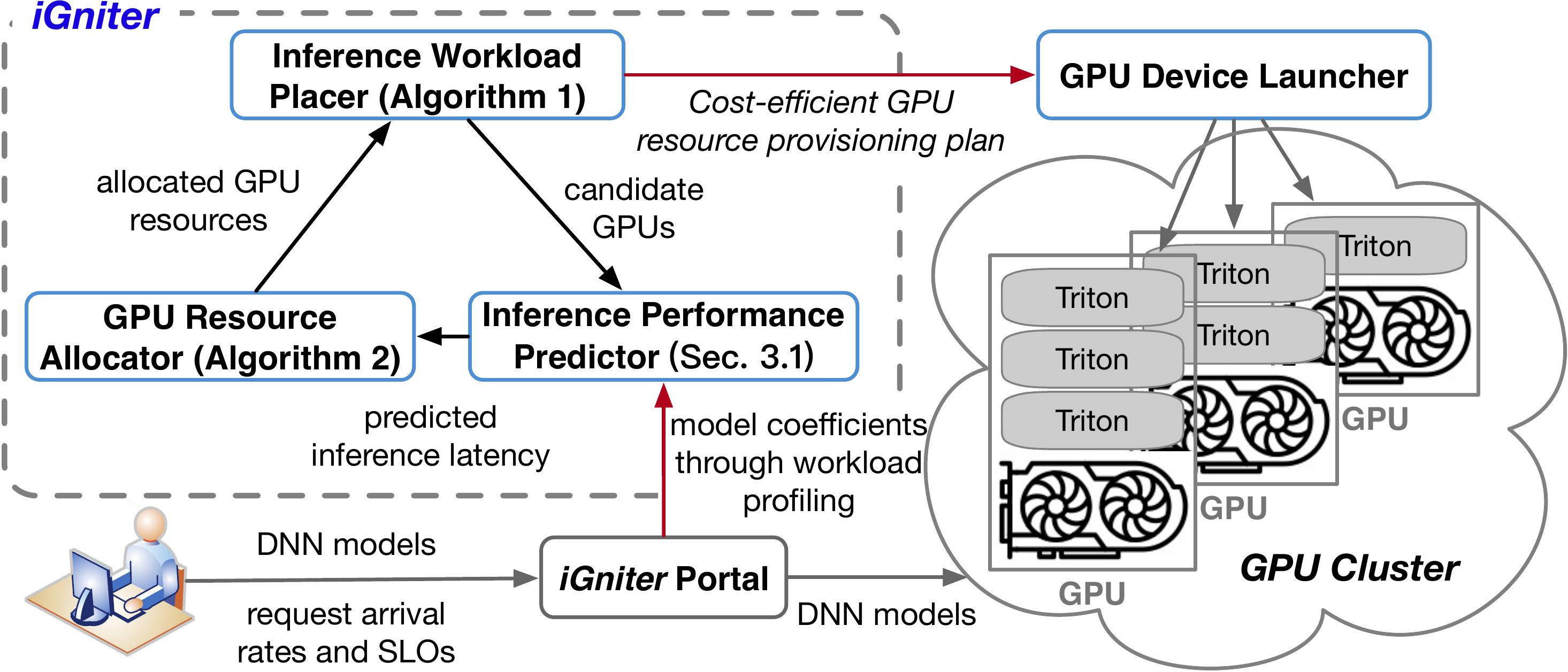}
	\caption{Overview of our \emph{iGniter} prototype in a GPU cluster.}
	\label{fig-prototype}\vspace{-10pt}
\end{figure}

\emph{iGniter} is \emph{periodically executed} to provision GPU resources for newly-arrived inference workloads. As illustrated in Fig.~\ref{fig-prototype}, \emph{iGniter} comprises three pieces of modules: an \emph{inference workload placer} and a \emph{GPU resource allocator} as well as an \emph{inference performance predictor}. Specifically, users submit DNN models with their request arrival rates and SLOs to the \emph{iGniter portal}, which can be deployed on a low-end EC2 instance. It initiates a \emph{lightweight} workload profiling on \emph{different types} of GPU devices to acquire the workload-specific and hardware-specific coefficients as elaborated in Sec.~\ref{sec:model-performance}. With such coefficients, the \emph{inference performance predictor} first estimates the inference latency using our performance model designed in Sec.~\ref{sec:model-performance}. It then guides our \emph{GPU resource allocator} and \emph{inference workload placer} to identify an \emph{appropriate} GPU device with the \emph{least} performance interference and guaranteed SLOs from candidate GPUs. To particularly offset the interference impact, Alg.~\ref{alg-resource-allocation} can judiciously adjust allocated GPU resources for both the newly-arrived and originally-placed inference workloads on a GPU device. According to our cost-efficient GPU resource provisioning plan generated by Alg.~\ref{alg-inference-provisioning}, the \emph{GPU device launcher} finally builds a GPU cluster and launches the Triton inference serving process for each DNN inference workload on the provisioned GPU devices. In particular, the inference batch size is configured in Triton, and the GPU resources are allocated to each Triton process using the $\mathtt{set\_active\_thread\_percentage}$ command in MPS.

\textbf{Dealing with Performance Prediction Errors.} The performance prediction errors can cause GPU resource \emph{under-provisioning} to DNN inference workloads, thereby resulting in SLO violations. \emph{iGniter} deals with such violations simply by pre-launching a \emph{shadow} Triton inference serving process \emph{standby} for each workload on a GPU device. Compared with the \emph{original} inference process, such a \emph{shadow} process is allocated an \emph{extra} amount of GPU resources when \emph{active}, which is set as the smaller value of the $10.0\%$ of GPU resources (\emph{i.e.,} the maximum prediction error measured in Sec.~\ref{sec:eval-prediction}) and the remaining resources on a GPU device. Specifically, the DNN inference requests are first sent to the original Triton inference serving process. User clients then continuously monitor the accumulated P99 latency of each inference workload every second. Once the P99 latency of inference requests violates the latency SLO, \emph{iGniter activates} the \emph{shadow} inference process and kills the original process. It then \emph{redirects} the upcoming inference requests to the \emph{activated shadow} process. We will validate the robustness of \emph{iGniter} in handling the performance prediction errors of DNN inference workloads in Sec.~\ref{sec:eval-effectiveness}.

\section{Performance Evaluation}
\label{sec:evaluation}

In this section, we evaluate \emph{iGniter} by carrying out a set of prototype experiments with four representative DNN models (as listed in Table~\ref{table-workload}) on Amazon EC2~\cite{ec2}. Our prototype experiments seek to answer the following questions:
\begin{itemize}
  \item \textbf{Accuracy:} Can our inference performance model in \emph{iGniter} accurately predict the performance of DNN inference workloads? (Sec.~\ref{sec:eval-prediction})
  \item \textbf{Effectiveness:} Can our GPU resource provisioning strategy in \emph{iGniter} provide predictable DNN inference while saving the monetary cost in the cloud? (Sec.~\ref{sec:eval-effectiveness})
  \item \textbf{Overhead:} How much runtime overhead of workload profiling and algorithm computation does \emph{iGniter} practically bring? (Sec.~\ref{sec:eval-overhead})
\end{itemize}

\subsection{Experimental Setup}
\label{sec:eval-experimental}

\textbf{GPU Cluster Configurations.} We set up a GPU cluster of $10$ p3.2xlarge EC2 instances, each equipped with $1$ NVIDIA V100 GPU card, $8$ vCPUs, and $61$ GB memory. On each instance, we launch a Triton inference serving process and its corresponding client with a constant request arrival rate for each DNN inference workload. We measure the seven \emph{hardware-specific} coefficients using the $\mathtt{Nsight}$ $\mathtt{Systems}$ and $\mathtt{nvidia-smi}$ according to Sec.~\ref{sec:model-performance}. The maximum power $P$, maximum frequency $F$, idle power $p_{idle}$, and available PCIe bandwidth $B_{pcie}$ of NVIDIA V100 are $300$ W, $1530$ MHz, $53.5$ W, and $10$ GBps, respectively. The power coefficient $\alpha_{f}$, scheduling coefficients $\alpha_{sch}$ and $\beta_{sch}$ are profiled as $-1.025$, $0.00475$ and $-0.00902$, respectively.

\textbf{Configurations of DNN Inference Workloads.} We select four representative DNN models as listed in Table~\ref{table-workload}. The AlexNet~\cite{AIG2017}, ResNet-50~\cite{KXSJ2016}, and VGG-19~\cite{KA2015} models are used for image classification running on the ImageNet dataset~\cite{JWRLKL2009}, while the SSD~\cite{WDDCSCA2016} model is used for object detection running on the VOC2012 dataset~\cite{MJ2012}. The four models (AlexNet, ResNet-50, VGG-19, and SSD) have \emph{heterogeneous} workload characteristics, \emph{i.e.,} computation complexity (GFLOPs) and model size (parameters), as elaborated in Table~\ref{table-workload}. In particular, we use \{$W1$, $\cdots$, $W12$\} to denote the $12$ DNN inference workloads with various performance SLOs in terms of latency SLOs and request arrival rates (\emph{i.e.,} expected throughputs) for App$1$, App$2$, and App$3$.

\begin{table}[!t]\vspace{+6pt}
\renewcommand{\arraystretch}{1.3}
\centering
\caption{Configurations of three Apps with four performance SLOs, \emph{i.e.,} latency (ms) and throughput (req/s) for four representative DNN inference models with \emph{heterogeneous} workload characteristics.}
\label{table-workload}
\begin{tabular}{c|c|cccc}
\toprule[1pt]
\multicolumn{2}{c|}{Workload features} & AlexNet & ResNet-50 & VGG-19 & SSD\\
\midrule[1pt]
\multicolumn{2}{c|}{GFLOPs} & $0.77$ & $4.14$ & $19.77$ & $62.82$ \\
\hline
\multicolumn{2}{c|}{Params (MB)} & $61.10$ & $25.56$ & $143.67$ & $26.29$ \\
\midrule[1pt]
\multirow{2}{*}{\rotatebox{90}{App$1$}} & Latency & $10$ & $20$ & $20$ & $25$ \\
	\cline{2-6}
	& Throughput & $1200$ & $400$ & $300$ & $150$\\
\hline
\multirow{2}{*}{\rotatebox{90}{App$2$}} & Latency & $15$ & $30$ & $30$ & $40$ \\
	\cline{2-6}
	& Throughput & $400$ & $600$ & $400$ & $50$\\
\hline
\multirow{2}{*}{\rotatebox{90}{App$3$}} & Latency & $20$ & $40$ & $40$ & $55$ \\
	\cline{2-6}
	& Throughput & $800$ & $200$ & $200$ & $300$\\
\bottomrule[1pt]
\end{tabular}\vspace{-10pt}
\end{table}

\begin{figure*}
	\begin{minipage}[t]{0.32\linewidth}
		\centering
		\includegraphics[width=2.27in]{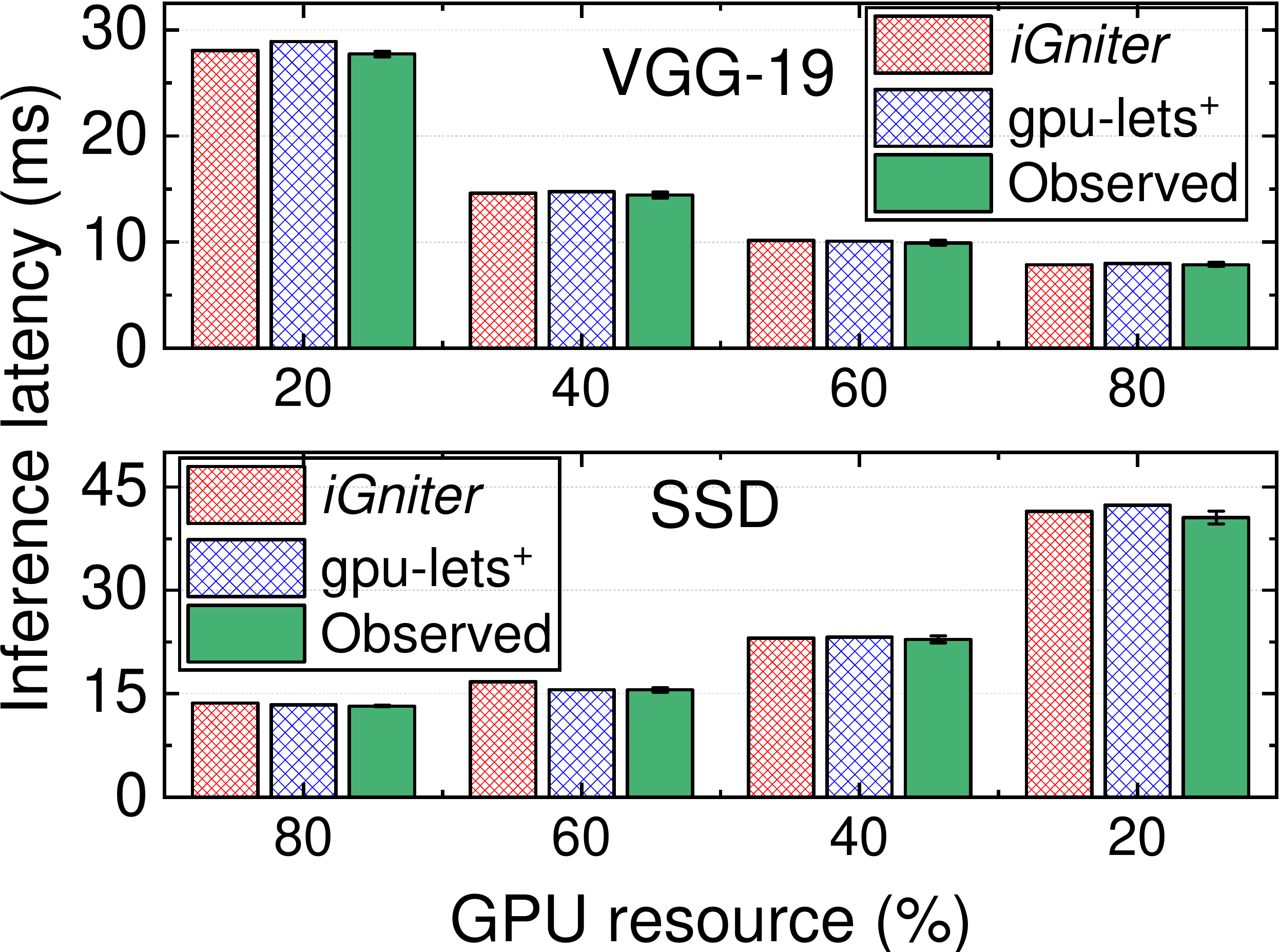}
		\caption{Comparison of the observed and predicted inference latency of co-located VGG-19 and SSD with different allocated GPU resources and batch size set as $3$ under the gpu-lets$^{+}$ and \emph{iGniter} performance models.}
		\label{fig-evaluation-predict-resource}
	\end{minipage}\hspace{+6pt}
	\begin{minipage}[t]{0.32\linewidth}
		\centering
		\includegraphics[width=2.3in]{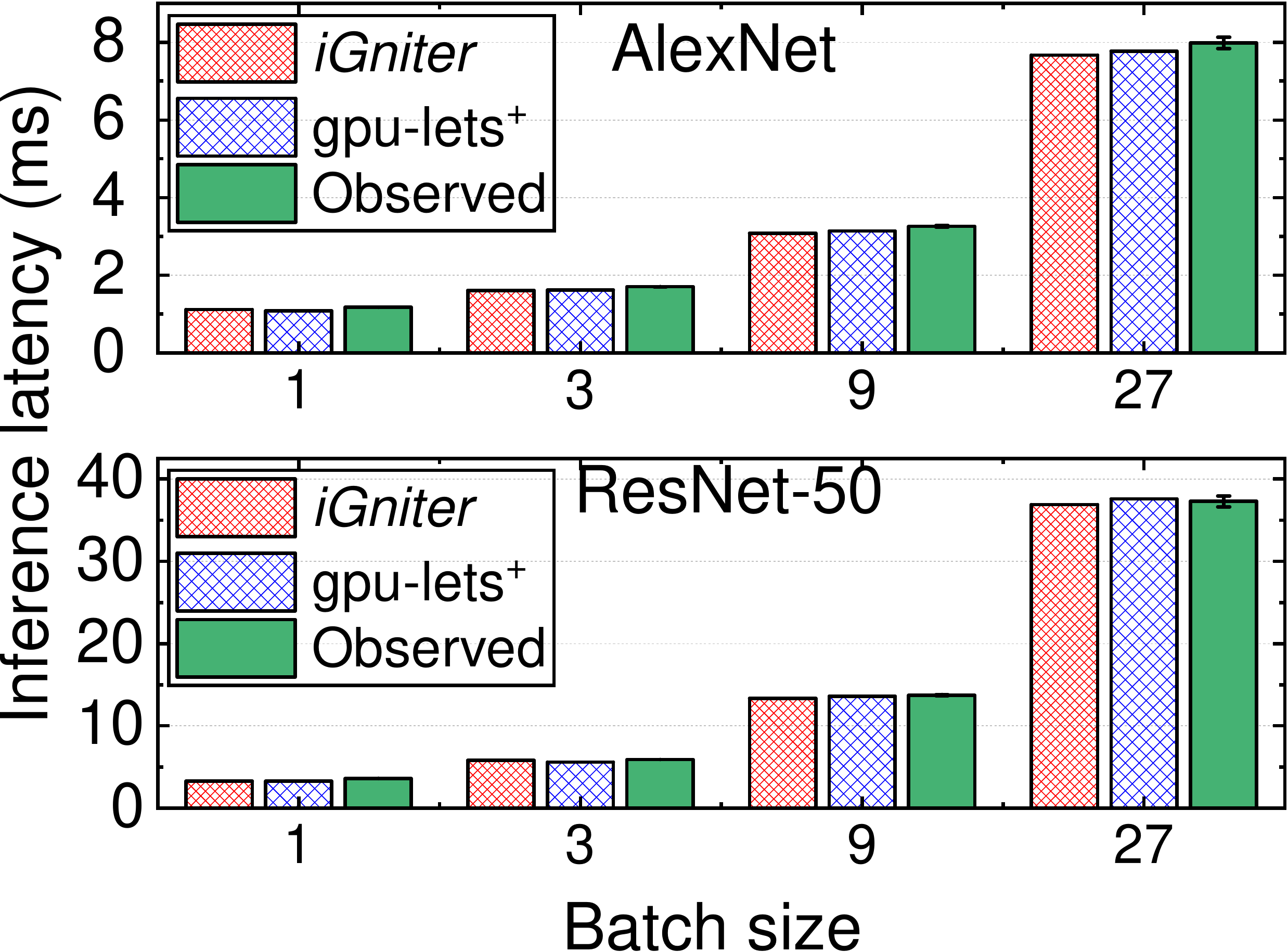}
		\caption{Comparison of the observed and predicted inference latency of co-located AlexNet and ResNet-50 with $50\%$ of allocated GPU resources and different batch sizes under the gpu-lets$^{+}$ and \emph{iGniter} performance models.}
		\label{fig-evaluation-predict-batch}
	\end{minipage}\hspace{+6pt}
	\begin{minipage}[t]{0.32\linewidth}
		\centering
		\includegraphics[width=2.2in]{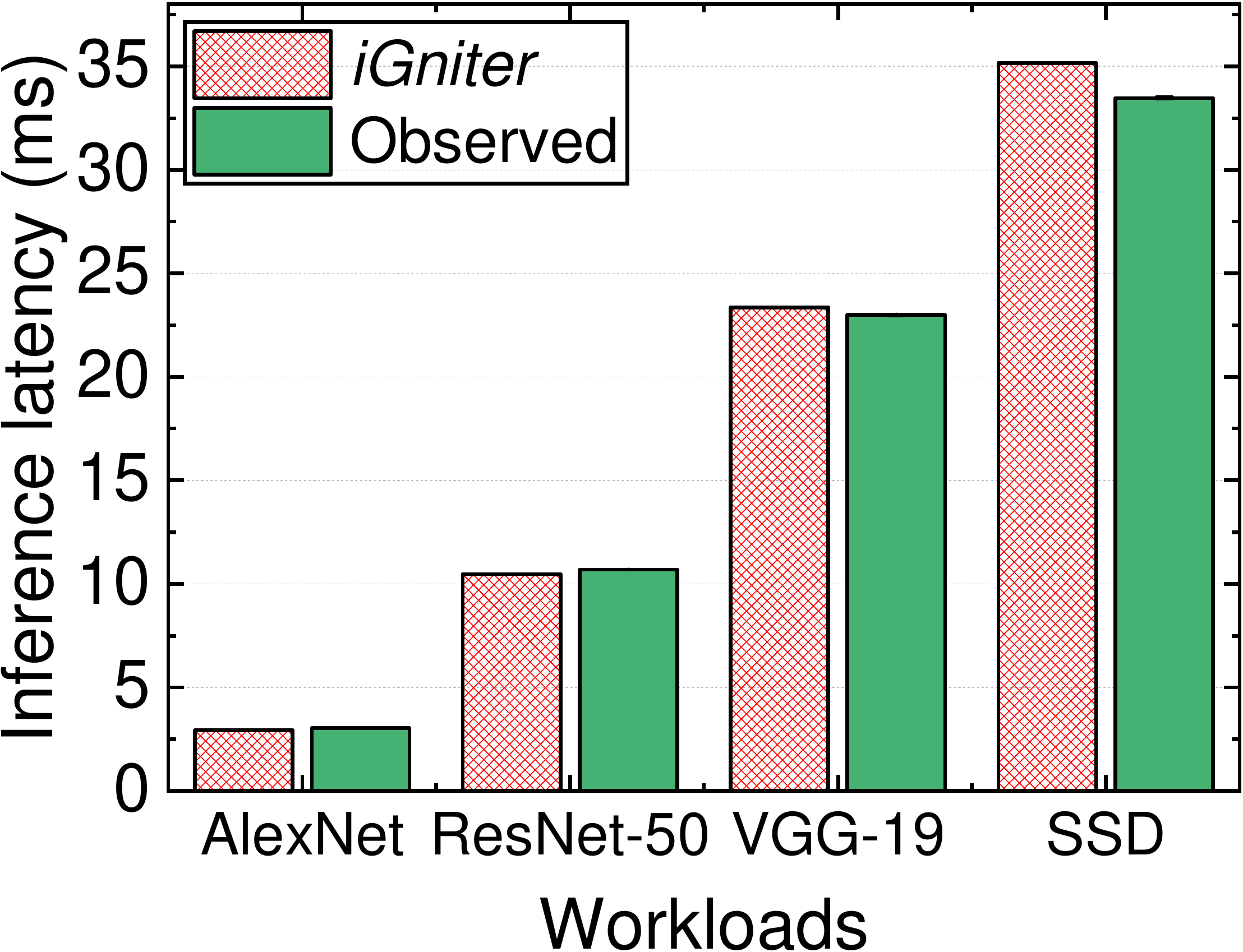}
		\caption{Comparison of the observed and \emph{iGniter} predicted inference latency of co-located AlexNet, ResNet-50, VGG-19 and SSD with $25\%$ of allocated GPU resources and batch size set as $3$.}
		\label{fig-evaluation-four-workloads}	
	\end{minipage}\vspace{-10pt}
\end{figure*}

\textbf{Baselines and Metrics.} We compare \emph{iGniter} with the following three strategies: (1) \textbf{FFD$^{+}$:} the First-Fit Decreasing (FFD) algorithm which always allocates the lower bound of GPU resources $r_{lower}^{i}$ and places inference workloads using FFD; (2) \textbf{GSLICE$^{+}$:} GSLICE~\cite{ASK2020} patched with our inference workload placement strategy, which tunes the allocated GPU resources and batch sizes according to the average latency and throughput of workloads; (3) \textbf{gpu-lets$^{+}$:} the modified gpu-lets~\cite{SSYJYJ2022}, which allocates the GPU resources by maximizing the request throughput and places inference workloads on the best-fit GPUs. We also change the batch size configuration strategy of gpu-lets$^{+}$ by increasing the batch size to \emph{just} meet the request arrival rate (the same as \emph{iGniter}), as large batch sizes cannot adapt to a low request arrival rate as evidenced in Sec.~\ref{sec:motivation-example}. In addition, we focus on two key metrics including the \emph{monetary cost} and \emph{SLO violations}, as elaborated in Sec.~\ref{sec:motivation-example}. We particularly calculate the \emph{hourly} monetary cost ($\$/h$) by multiplying the number of provisioned GPU instances and the hourly price of each instance. We do not multiply it by the inference execution time, simply because the model inference requests arrive constantly from users in our scenario.

\subsection{Validating Inference Performance Model in \emph{iGniter}}
\label{sec:eval-prediction}

We evaluate the inference latency of AlexNet, ResNet-50, VGG-19, and SSD by varying the amount of GPU resources, batch size, and the number of co-located inference workloads. We compare our \emph{iGniter} performance model with the state-of-the-art gpu-lets$^{+}$ model~\cite{SSYJYJ2022}. We illustrate the observed inference latency with error bars of standard deviation by repeating experiments three times.

\textbf{Can \emph{iGniter} accurately predict the inference latency with different amounts of GPU resources?} As shown in Fig.~\ref{fig-evaluation-predict-resource}, \emph{iGniter} can well predict the inference latency with a prediction error of $0.04\%$ -- $2.32\%$ for VGG-19 and $0.89\%$ -- $7.61\%$ for SSD, compared with $1.30\%$ -- $4.19\%$ and $0.02\%$ -- $4.43\%$ under gpu-lets$^{+}$. Specifically, our predicted inference latency of SSD is basically higher than gpu-lets$^{+}$ and the observed latency. This is because the active time of SSD predicted by our model is longer than the actual active time, and the contention of GPU power consumption and L2 cache utilization further makes it worse. However, gpu-lets$^{+}$ offline profiles the actual inference latency for all possible configurations when SSD is running alone. In addition, the predicted inference latency of VGG-19 under \emph{iGniter} is more accurate than that under gpu-lets$^{+}$. This is because gpu-lets$^{+}$ does not consider the contention of the GPU scheduler and power consumption. The GPU frequency for running VGG-19 drops from $1,530$ MHz to $1,440$ MHz due to GPU power contention, which makes the prediction error of gpu-lets$^{+}$ larger than \emph{iGniter} for VGG-19.

\textbf{Can \emph{iGniter} accurately predict the inference latency with different batch sizes?} As depicted in Fig.~\ref{fig-evaluation-predict-batch}, \emph{iGniter} can basically predict the DNN inference latency with a prediction error of $3.91\%$ -- $5.90\%$ for AlexNet and $1.10\%$ -- $9.29\%$ for ResNet-50, compared with $2.67\%$ -- $6.23\%$ and $0.78\%$ -- $9.76\%$ of gpu-lets$^{+}$. Specifically, the predicted inference latency of AlexNet under \emph{iGniter} is smaller than the observed latency. This is because the data loading and result feedback phases occupy a larger part (\emph{i.e.,} $7\%$ -- $20\%$) of the inference latency for AlexNet than that for other models (\emph{i.e.,} $1\%$ -- $7\%$). It makes AlexNet share the PCIe bandwidth for a long period of time with other workloads. However, we simply assume that the contention of the PCIe bandwidth can be negligible. Also, \emph{iGniter} underestimates the inference latency of ResNet-50 with a prediction error of $9.29\%$ when the batch size is set as $1$. This is because the average GPU active time of ResNet-50 is relatively small (\emph{i.e.,} $0.04$ ms), which makes it more sensitive to the GPU scheduler contention than other workloads. As \emph{iGniter} explicitly considers such contention of GPU scheduler, the average prediction error of \emph{iGniter} (\emph{i.e.,} $3.82\%$) is smaller than that of gpu-lets$^{+}$ (\emph{i.e.,} $4.15\%$) for ResNet-50.

\textbf{Can \emph{iGniter} adapt to the co-location of multiple ($4+$) inference workloads?} As shown in Fig.~\ref{fig-evaluation-four-workloads}, we observe that \emph{iGniter} can accurately predict the inference latency of the four co-located workloads with a prediction error of $1.53\%$ -- $5.02\%$, while gpu-lets$^{+}$ fails to predict the inference latency of more than two co-located inference workloads. Specifically, our \emph{iGniter} model captures the interference on the GPU scheduler (Eq.~(\ref{eq-increased-schedule-latency})), L2 cache space (Eq.~(\ref{eq-cache-interference})), and power consumption (Eq.~(\ref{eq-frequency})) for multiple co-located inference workloads. Taking VGG-19 as an example, \emph{iGniter} can well predict the inference latency with a prediction error of $4.19\%$ when co-located only with SSD (in Fig.~\ref{fig-evaluation-predict-resource}) and $1.53\%$ when co-located with three inference workloads (\emph{i.e.,} AlexNet, ResNet-50, and SSD in Fig.~\ref{fig-evaluation-four-workloads}), respectively. The rationale is that: when VGG-19 is co-located with two more workloads (\emph{i.e.,} AlexNet, ResNet-50), \emph{iGniter} can still predict the increase of GPU scheduling delay from $0.19$ ms to $0.36$ ms and the decrease of GPU active time from $27.54$ ms to $22.31$ ms (as allocated $5\%$ more GPU resources), as well as the drop of GPU frequency from $1,530$ MHz to $1,515$ MHz.

%\begin{table}[!t]\vspace{+6pt}
%\renewcommand{\arraystretch}{1.3}
%\centering
%\caption{Comparison of the \emph{hourly} monetary cost and the number of P99 latency violations of inference workloads achieved by the gpu-lets$^{+}$, FFD$^{+}$, GSLICE$^{+}$, and \emph{iGniter} resource provisioning strategies.}
%\label{table-cost-violations}
%\begin{tabular}{c|cccc}
%\toprule[1pt]
%\multicolumn{2}{c}{\qquad \qquad \qquad \quad gpu-lets$^{+}$} & FFD$^{+}$ & GSLICE$^{+}$ & \emph{iGniter} \\
%\hline
%Cost ($\$/h$) & $24.48$ & $15.3$ & $18.36$ & $18.36$\\
%\hline
%Violations  & $3$ & $10$ & $3$ & $0$\\
%\bottomrule[1pt]
%\end{tabular}\vspace{-10pt}
%\end{table}

\begin{figure*}
	\centering
	\includegraphics[width=6.57in]{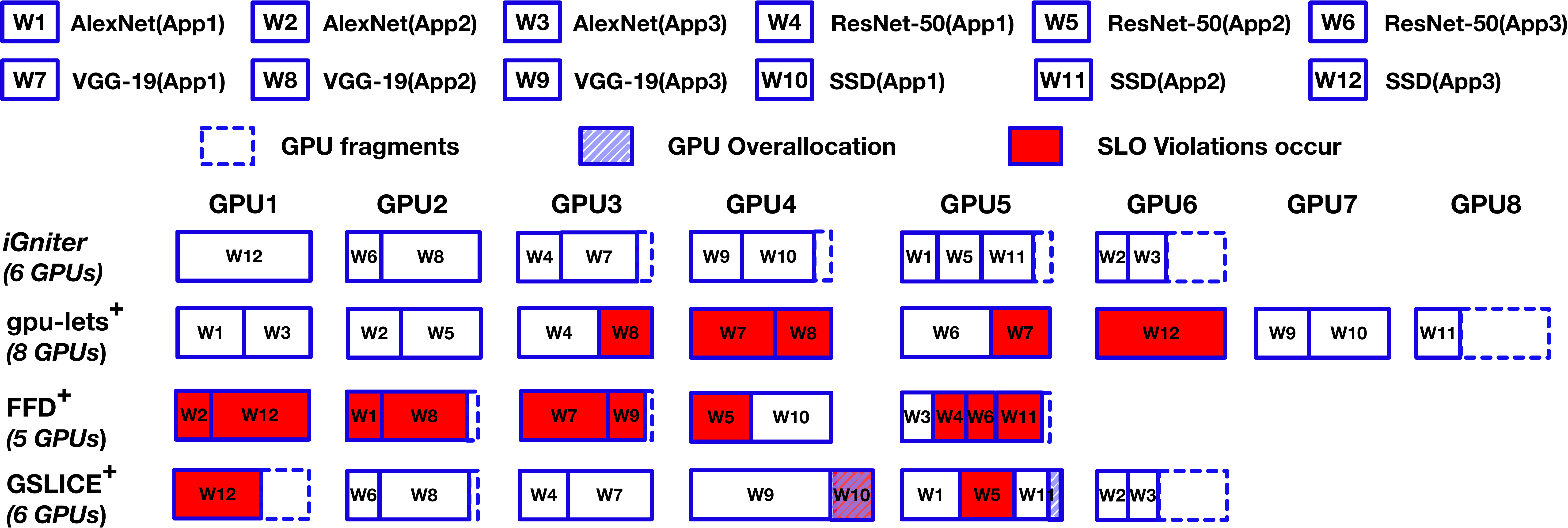}
	\caption{Comparison of GPU resource provisioning plans for the $12$ workloads (\emph{i.e.,} $W1$, $\cdots$, $W12$). \emph{iGniter}, gpu-lets$^{+}$, FFD$^{+}$, and GSLICE$^{+}$ provision $6$, $8$, $5$, and $6$ GPU devices (p3.2xlarge instances), which achieve $\$18.36$, $\$24.48$, $\$15.3$, and $\$18.36$ monetary cost per hour, respectively. In addition, the four GPU resource provisioning strategies bring $0$, $3$, $10$, and $3$ SLO violations, respectively.}
	\label{fig-evaluation-allocation}
	\vspace{-0pt}
\end{figure*}

\begin{figure*}
	\begin{minipage}[t]{0.32\linewidth}
		\centering
		\includegraphics[width=2.32in]{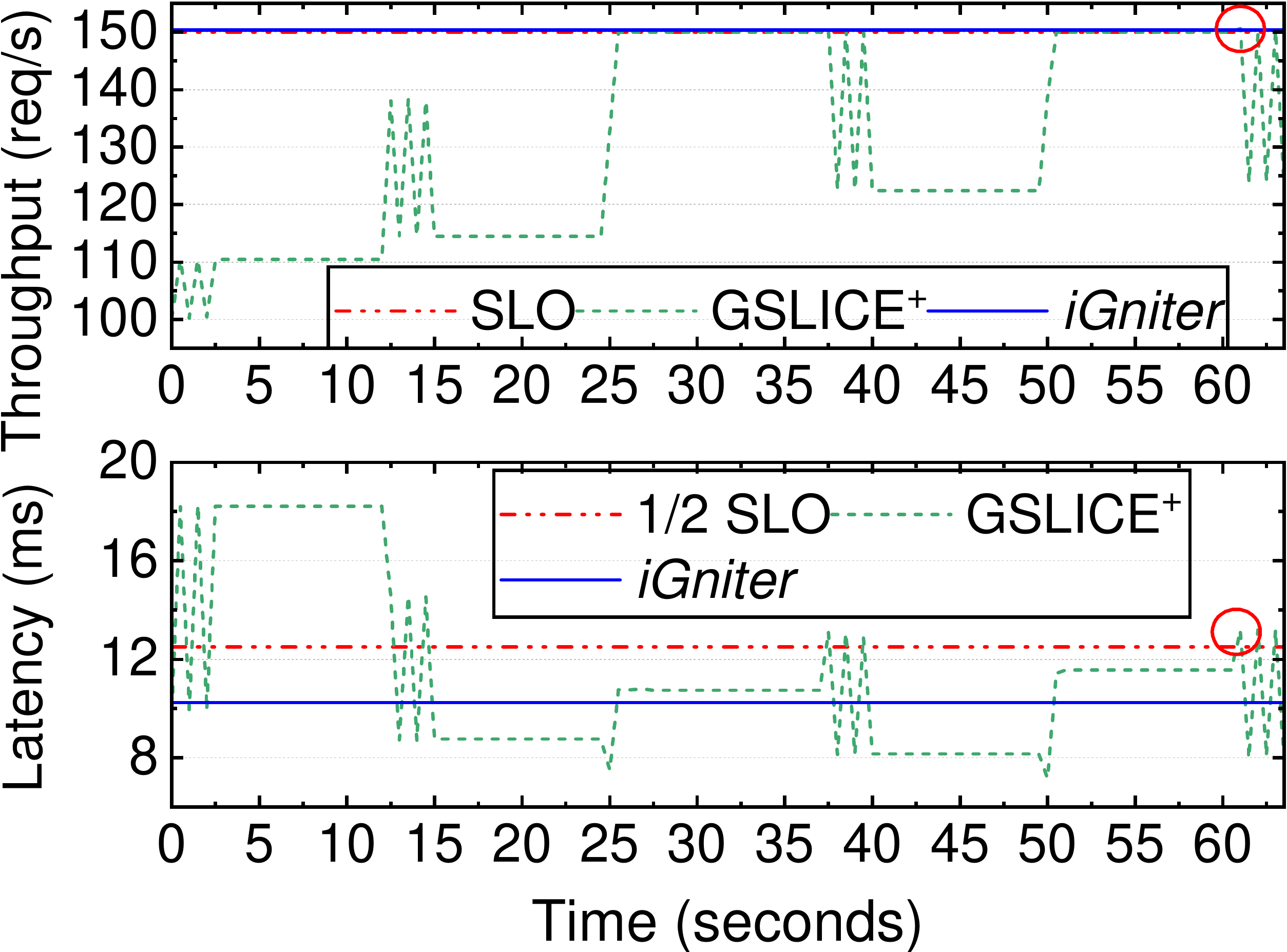}
		\caption{Comparison of the inference latency and request throughput of $W10$ over time under the GSLICE$^{+}$ and \emph{iGniter} strategies.}
		\label{fig-evaluation-gslice-performance}
	\end{minipage}\hspace{+8pt}
	\begin{minipage}[t]{0.32\linewidth}
		\centering
		\includegraphics[width=2.27in]{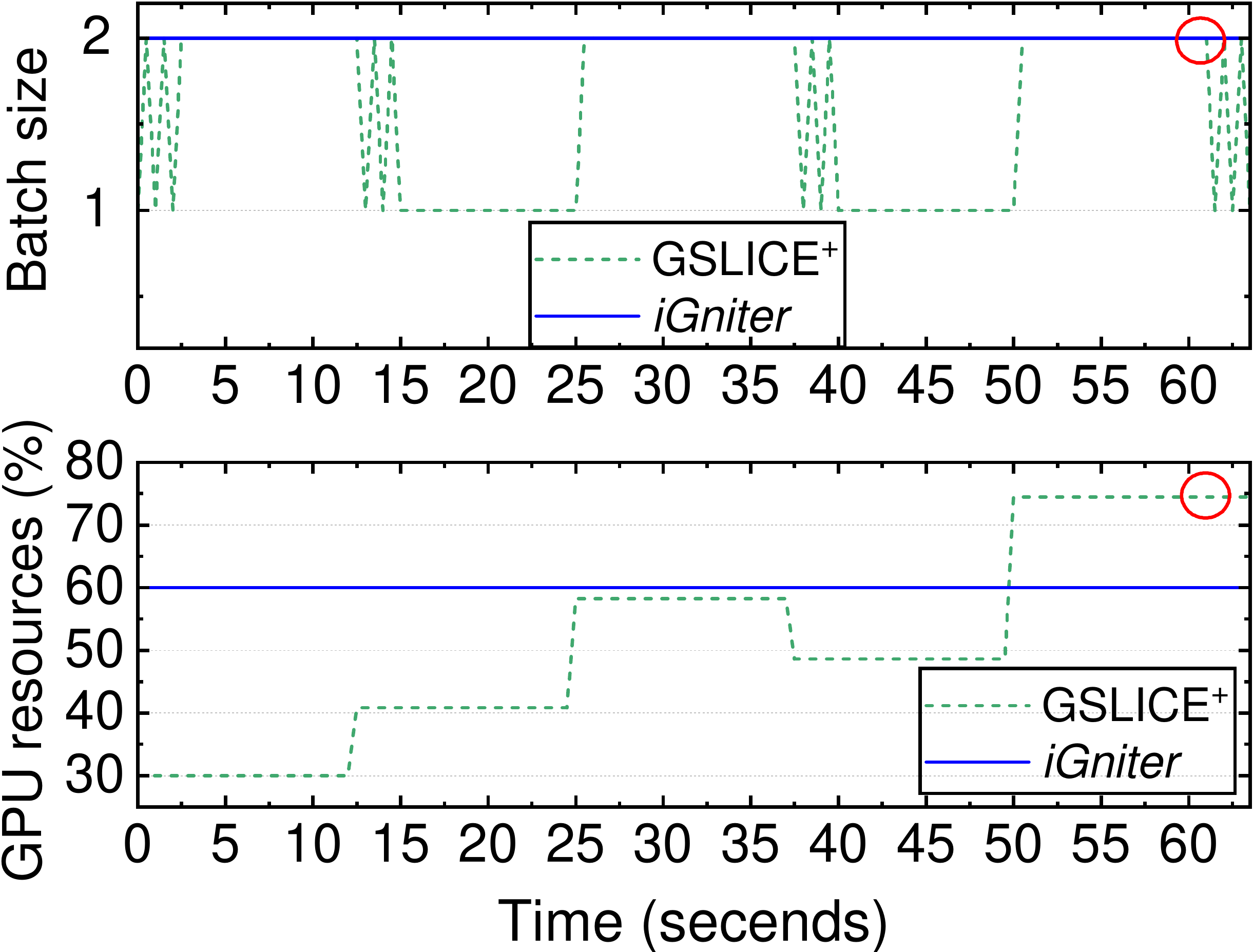}
		\caption{Comparison of the allocated GPU resources and batch sizes for $W10$ over time under the GSLICE$^{+}$ and \emph{iGniter} strategies.}
		\label{fig-evaluation-gslice-configuration}
	\end{minipage}\hspace{+3pt}
	\begin{minipage}[t]{0.32\linewidth}
		\centering
		\includegraphics[width=2.27in]{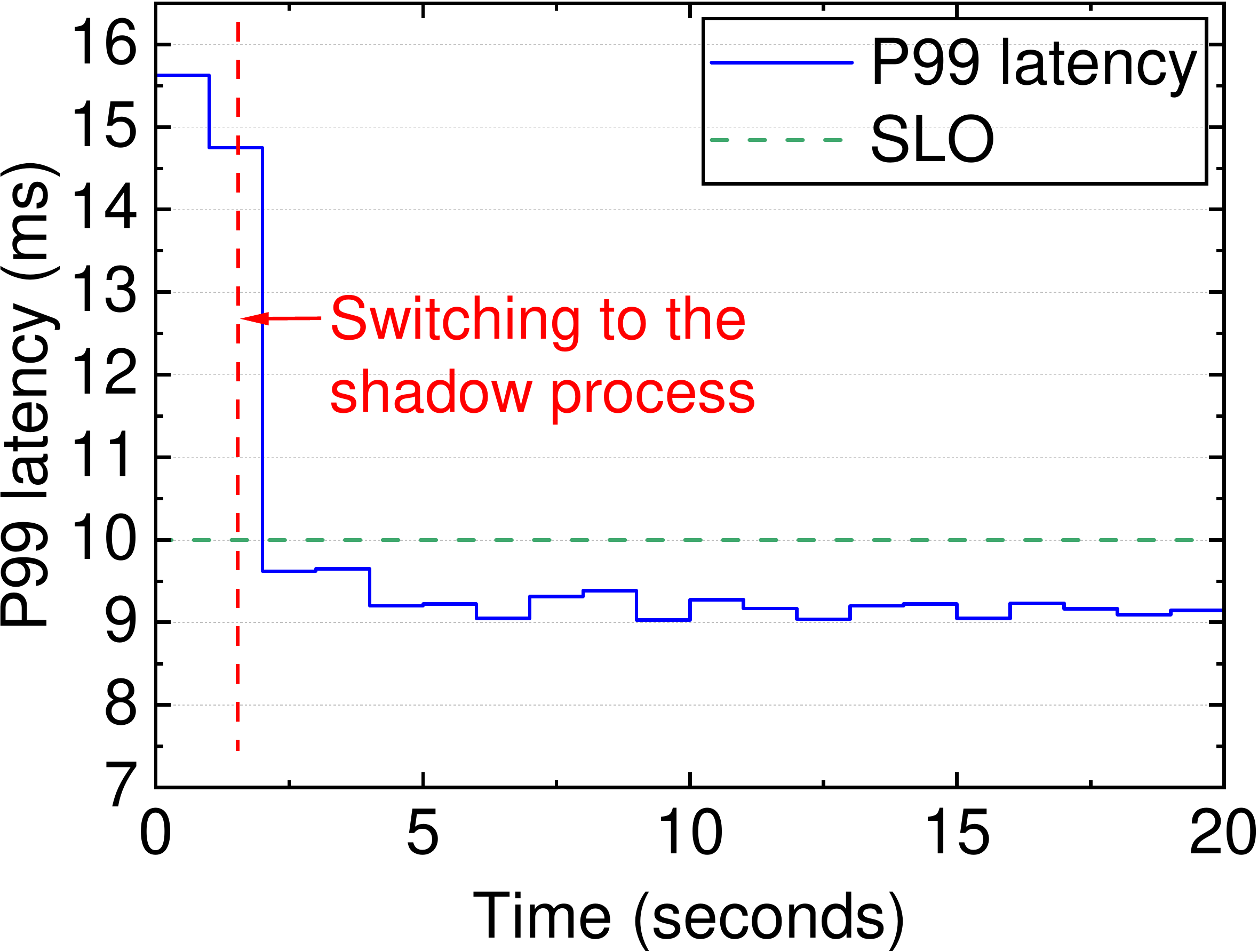}
		\caption{P99 inference latency of $W1$ (\emph{i.e.,} App$1$ of AlexNet) over time when \emph{iGniter} handles SLO violations.}
		\label{fig-evaluation-error-handling}	
	\end{minipage}\vspace{-10pt}
\end{figure*}

\subsection{Effectiveness of GPU Resource Provisioning Strategy in \emph{iGniter}}
\label{sec:eval-effectiveness}

To illustrate the effectiveness of our \emph{iGniter} resource provisioning strategy, we conduct extensive experiments with the $12$ inference workloads in Table~\ref{table-workload}. Specifically, we measure the P99 latency of inference workloads within a period of time (\emph{e.g.,} $30$ seconds). During the online resource adjustment, we adopt the resource provisioning plan after \emph{five} adjustments of GPU resources for GSLICE$^{+}$. Similarly, we select the resource provisioning plan \emph{after} dealing with prediction errors for \emph{iGniter}. As illustrated in Fig.~\ref{fig-evaluation-allocation}, \emph{iGniter} guarantees the P99 inference latency of all $12$ inference workloads within their latency SLOs, while saving up to $25\%$ of \emph{hourly} monetary cost compared with gpu-lets$^{+}$.

\textbf{How can \emph{iGniter} guarantee performance SLOs?} As shown in Fig.~\ref{fig-evaluation-allocation}, FFD$^{+}$ \emph{first} makes $10$ out of $12$ workloads violate performance SLOs because it does not consider the interference of co-located workloads. In contrast, \emph{iGniter} provisions an additional $25\%$ of GPU resources (\emph{i.e.,} GPU$6$) and adequately places workloads on GPUs to \emph{proactively} eliminate SLO violations caused by the interference. \emph{Second,} though gpu-lets$^{+}$ provisions the largest amount of GPU resources, there still exist $3$ workloads (\emph{i.e.,} $W7$, $W8$, $W12$) violating performance SLOs. This is because gpu-lets$^{+}$ does not model the interference on request throughputs and it simply uses the profiled throughput when the workload is running alone. It inevitably makes workloads easily violate the expected throughput. \emph{Third,} GSLICE$^{+}$ can cause $3$ violations even using our workload placement plan. This is because the \emph{interference-unaware} strategy (\emph{i.e.,} GSLICE$^{+}$) \emph{separately} adjusts allocated GPU resources and batch size according to a fixed tuning threshold (\emph{e.g.,} $10\%$), which can make the inference performance \emph{oscillate frequently} around SLOs. We take $W10$ (co-located with $W9$ on GPU$4$) as an example. As shown in Fig.~\ref{fig-evaluation-gslice-performance}, the average inference latency (\emph{i.e.,} $10.7$ ms) is lower than the $\frac{1}{2}$SLO (\emph{i.e.,} $12.5$ ms) exceeding the tuning threshold during $25.5$ -- $37.5$ seconds. It then triggers GSLICE$^{+}$ to reduce the allocated GPU resources, which makes SSD violate the expected throughput ($150$ req/s). Moreover, GSLICE$^{+}$ adjusts the GPU resources of $W9$ to $100\%$ at the $51$-th second without considering $W10$, and the resources are successfully allocated to $W9$ at the $61$-th second (\emph{i.e.,} the red circle in Fig.~\ref{fig-evaluation-gslice-performance} and Fig.~\ref{fig-evaluation-gslice-configuration}). In such a case, the overallocation of GPU resources occurs, which brings SLO violations to both $W9$ and $W10$. In contrast, \emph{iGniter} leverages our analytical inference performance model to \emph{proactively} provision an adequate amount of GPU resources and to configure an appropriate batch size when launching inference workloads on GPUs.

\textbf{Can \emph{iGniter} deal with the performance prediction errors?} The prediction error handling mechanism in \emph{iGniter} further guarantees performance SLOs. In our experiments, such a mechanism only triggers two times (\emph{i.e.,} two prediction errors occur). To illustrate how it works, we take $W1$ co-located with $W5$ and $W11$ on GPU$5$ as an example. As depicted in Fig.~\ref{fig-evaluation-error-handling}, the P99 latency of $W1$ at the first second is $15.6$ ms which is higher than the latency SLO (\emph{i.e.,} $10$ ms) due to the prediction error. In the next $0.5$ seconds, \emph{iGniter} collects the request latency data and judges whether it violates the SLO. If an SLO violation still occurs, \emph{iGniter} switches such an SLO-violated inference workload to the \emph{activated} shadow Triton process at the $1.5$-th second. After that, the P99 latency of $W1$ can be guaranteed within the SLO. As we have pre-launched the \emph{shadow} Triton process as elaborated in Sec.~\ref{sec:design-implement}, \emph{iGniter} does not require spending $10$ seconds in launching a new Triton process as in GSLICE$^{+}$.

\begin{figure}[!t]
\centering\includegraphics[width=3.5in]{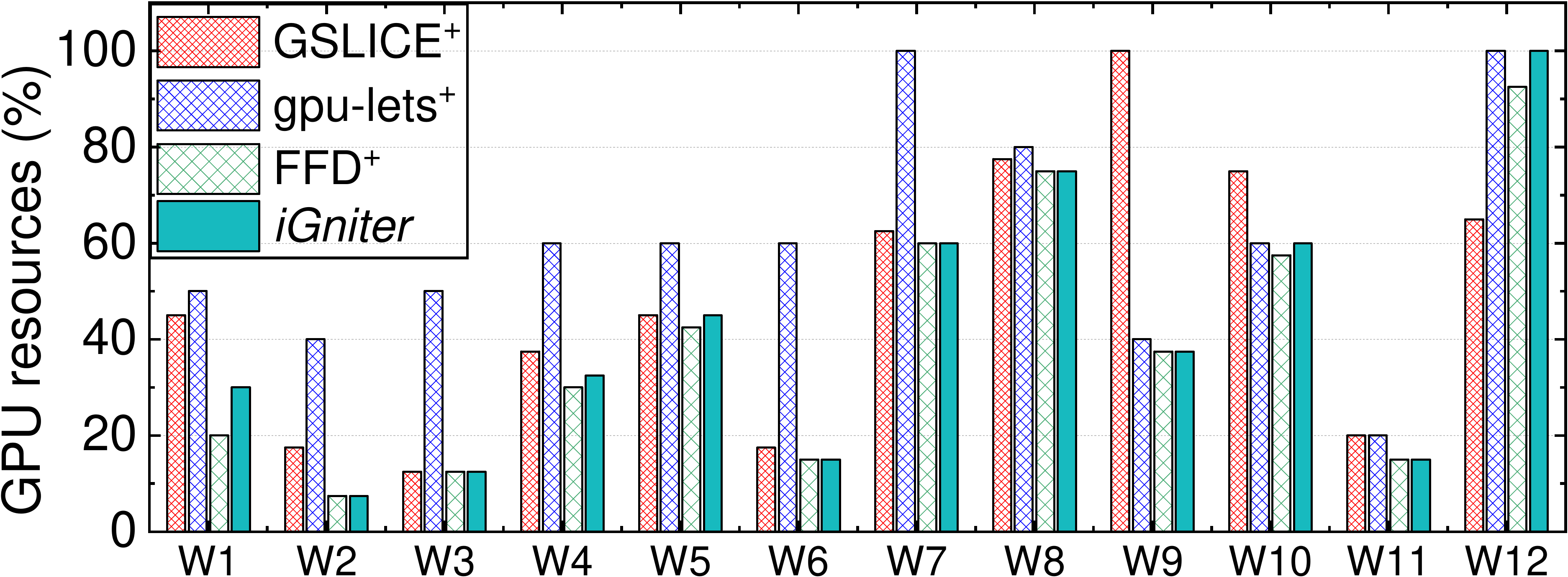}
\caption{Comparison of allocated GPU resources for the $12$ workloads (\emph{i.e.,} $W1$, $\cdots$, $W12$) achieved by the gpu-lets$^{+}$, FFD$^{+}$, GSLICE$^{+}$, and \emph{iGniter} strategies.}
\label{fig-evaluation-provisioned-resources}\vspace{-0pt}
\end{figure}

\textbf{How can \emph{iGniter} save the monetary cost?} As the \emph{hourly} monetary cost is proportional to the number of provisioned GPU instances, we simply compare the \emph{allocated GPU resources} of \emph{iGniter} with that of GSLICE$^{+}$, FFD$^{+}$, and gpu-lets$^{+}$. As shown in Fig.~\ref{fig-evaluation-provisioned-resources}, we observe that the GPU resources allocated by gpu-lets$^{+}$ for each workload are larger or equal to \emph{iGniter}. This is mainly due to the following facts: \emph{First,} taking $W4$ (\emph{i.e.,} App$1$ of ResNet-50) as an example, gpu-lets$^{+}$ provisions $60\%$ of GPU resources (\emph{i.e.,} the most-efficient amount of GPU resources) and then sets the batch size as $2$ to maximize its throughput. In contrast, \emph{iGniter} sets an \emph{appropriate} batch size as $4$ and then provisions $32.5\%$ of GPU resources to just meet its performance SLOs. \emph{Second,} gpu-lets$^{+}$ only allows two co-located inference workloads on a GPU device, while \emph{iGniter} allows multiple (more than $2$) workloads concurrently executed. \emph{Third,} gpu-lets$^{+}$ allows only \emph{five} choices (\emph{i.e.,} $20\%$, $40\%$, $50\%$, $60\%$, $80\%$) of GPU resources allocated to inference workloads, while \emph{iGniter} can allocate workloads with an amount of GPU resources with a \emph{fine-grained} GPU allocation unit (\emph{i.e.,} $2.5\%$). For example, gpu-lets$^{+}$ and \emph{iGniter} provision $W9$ with $40\%$ and $37.5\%$ of GPU resources, respectively. In addition, though GSLICE$^{+}$ uses our workload placement plan, it provisions more or equal amounts of GPU resources than \emph{iGniter} for all workloads except $W12$ which violates its latency SLO. This is because GSLICE$^{+}$ does not reduce its allocated GPU resources, as long as an inference workload meets its performance SLOs and the tuning threshold. FFD$^{+}$ provisions less or equal amounts of GPU resources than \emph{iGniter} as it always allocates the lower bound ($r_{lower}^{i}$) of GPU resources to inference workloads.

\begin{figure}[!t]
	\centering\includegraphics[width=3.5in]{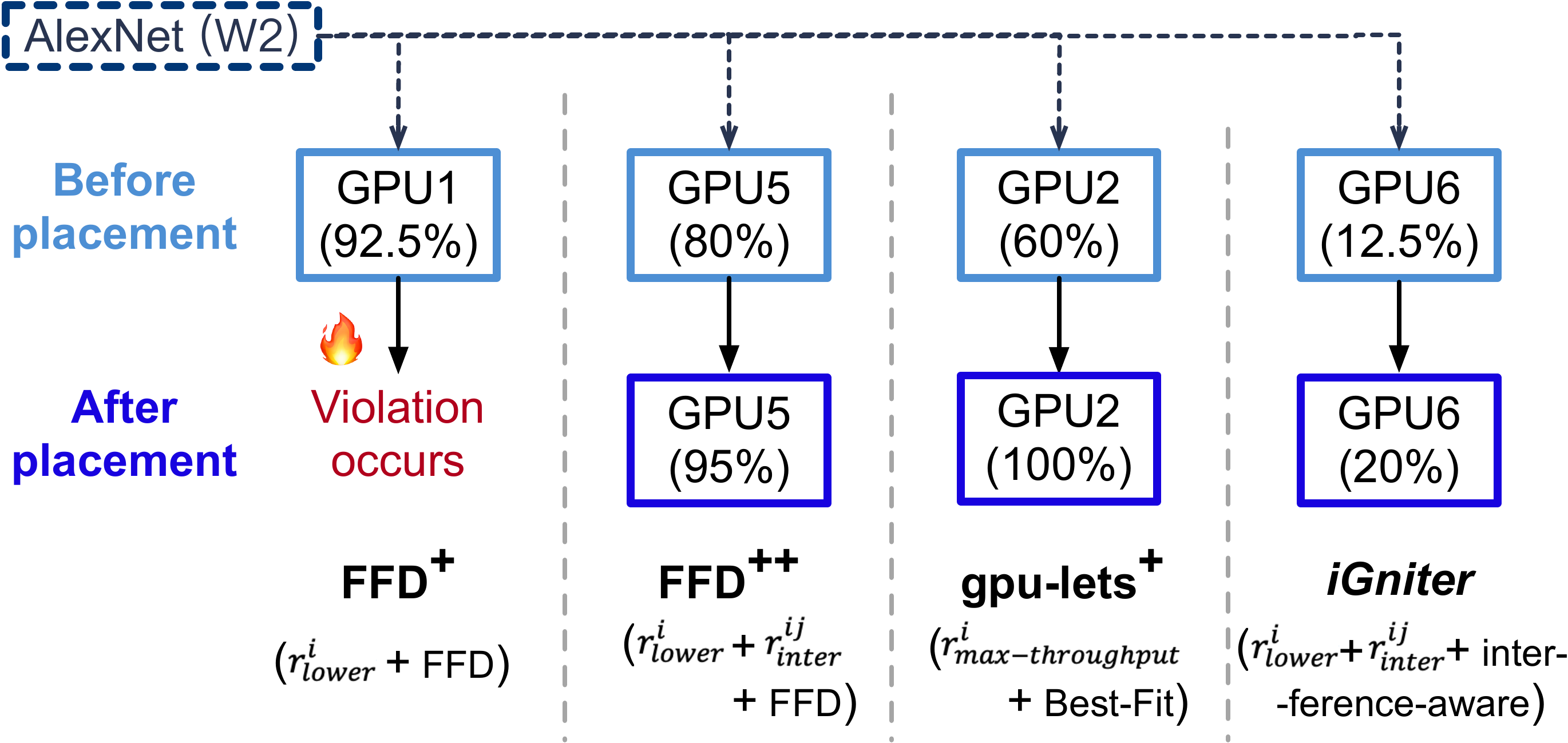}
	\caption{Comparison of the inference workload (\emph{i.e.,} App$2$ of AlexNet) placement decisions achieved by the FFD$^{+}$, gpu-lets$^{+}$, FFD$^{++}$ (\emph{i.e.,} FFD$^{+}$ using $\mathtt{alloc\_gpus}$, Alg.~\ref{alg-resource-allocation}), and \emph{iGniter} resource provisioning strategies.}
	\label{fig-evaluation-placement}\vspace{-10pt}
\end{figure}

\textbf{How can \emph{iGniter} place inference workloads on GPUs?} The \emph{inference workload placer} elaborated in Sec.~\ref{sec:design-implement} in \emph{iGniter} further reduces the amount of allocated GPU resources. As shown in Fig.~\ref{fig-evaluation-placement}, FFD$^{+}$ places $W2$ (\emph{i.e.,} App$2$ of AlexNet) onto GPU$1$ according to the lower bound of GPU resources (\emph{i.e.,} $r_{lower}^{i}$) which inevitably causes SLO violations due to the overlooked performance interference. FFD$^{++}$ places such a workload onto GPU$5$ with $15\%$ of GPU resources according to the first-fit GPU that still has an amount (\emph{i.e.,} $r_{lower}^{i} + r_{inter}^{ij}$ which is calculated by Alg.~\ref{alg-resource-allocation}) of GPU resources. As the most-efficient amount of GPU resources (\emph{i.e.,} $r_{max\_throughput}^{i}$) for App$2$ of AlexNet is $40\%$, gpu-lets$^{+}$ places $W2$ onto GPU$2$ which is selected as the best-fit GPU device. In general, gpu-lets$^{+}$ allocates more GPU resources than the other strategies as it mainly focuses on improving the inference throughput. In contrast, \emph{iGniter} places $W2$ onto GPU$6$ with the least amount of GPU resources ($7.5\%$) while guaranteeing the latency SLOs of all workloads. This is because \emph{iGniter} greedily places the inference workload onto the GPU with the least performance interference and allocates GPU resources that \emph{just meet} performance SLOs.

\begin{figure}[!t]
	\centering\includegraphics[width=3.5in]{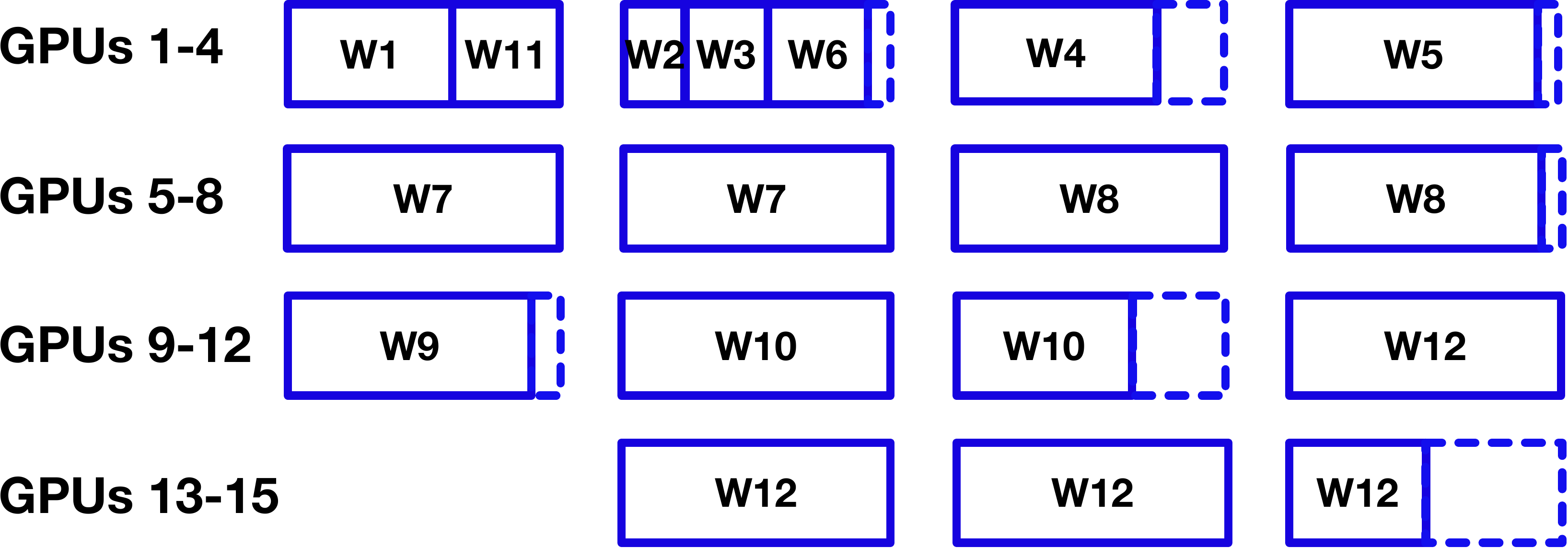}
	\caption{GPU resource provisioning plans achieved by \emph{iGniter} for the $12$ workloads in a cluster of $15$ g4dn.xlarge instances without any SLO violations, resulting in $\$7.89$ monetary cost per hour.}
	\label{fig-evaluation-T4-allocation}\vspace{-0pt}
\end{figure}

\textbf{Can \emph{iGniter} adapt to the heterogeneous cluster?} To obtain complementary insights, we extend our GPU cluster by adding $20$ g4dn.xlarge instances, each equipped with $1$ NVIDIA T4 GPU card, $4$ vCPUs, and $16$ GB memory. After obtaining the \emph{hardware-specific} coefficients and a part of \emph{workload-specific} coefficients on the g4dn.xlarge instance, Alg.~\ref{alg-inference-provisioning} can identify the appropriate GPU resource provisioning plan as illustrated in Fig.~\ref{fig-evaluation-T4-allocation}. As the NVIDIA V100 GPU device is equipped with $2 \times$ GPU computing resources and $3 \times$ memory bandwidth resources compared with the NVIDIA T4 GPU device, \emph{iGniter} provisions $15$ g4dn.xlarge instances (T4) while $6$ p3.2xlarge instances (V100) for the $12$ workloads, respectively. In particular, \emph{iGniter} provisions $2+$ g4dn.xlarge instances for $W7$, $W8$, $W10$, and $W12$ to meet their performance SLOs. Finally, as the hourly monetary cost (\emph{i.e.,} $\$7.89$) on g4dn.xlarge instances is much less than that (\emph{i.e.,} $\$18.36$) on p3.2xlarge instances, \emph{iGniter} considers g4dn.xlarge as the most cost-efficient type of instances and it adopts the resource provisioning plan in Fig.~\ref{fig-evaluation-allocation} for serving the $12$ inference workloads.

\begin{figure}[!t]
	\centering\includegraphics[width=2.6in]{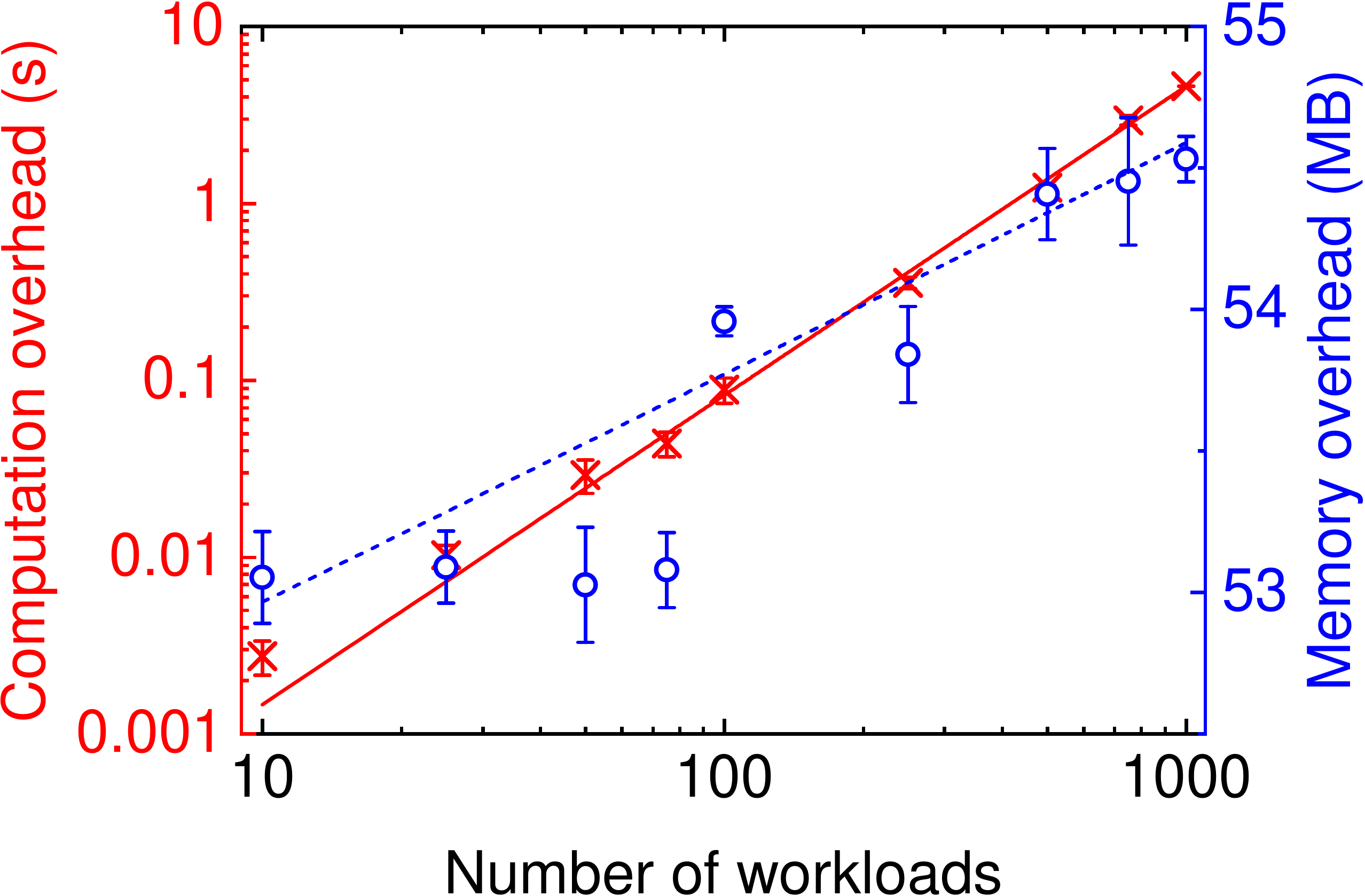}
	\caption{Computation and memory overhead of \emph{iGniter} by varying the number of DNN inference workloads from $10$ to $1,000$.}
	\label{fig-evaluation-scalability}\vspace{-10pt}
\end{figure}

\subsection{Runtime Overhead of \emph{iGniter}}
\label{sec:eval-overhead}

We evaluate the runtime overhead of \emph{iGniter} in terms of the profiling overhead of DNN inference workloads, and the computation time and memory consumption of \emph{iGniter} resource provisioning strategy (\emph{i.e.,} Alg.~\ref{alg-inference-provisioning}). Specifically, we launch a p3.2xlarge EC2 instance to profile the \emph{workload-specific} coefficients \emph{only once} for each inference workload. The profiling time of AlexNet~\cite{AIG2017}, ResNet-50~\cite{KXSJ2016}, VGG-19~\cite{KA2015}, and SSD~\cite{WDDCSCA2016} models are $231$, $247$, $240$, and $237$ seconds, respectively. In addition, we profile the \emph{hardware-specific} coefficients with VGG-19 \emph{only once for a given GPU type} and the profiling time is merely $229$ seconds. The experiment results above show that the profiling overhead of inference workloads is within several (around $4$) minutes, which is far less than the runtime overhead of gpu-lets~\cite{SSYJYJ2022} (\emph{i.e.,} over several hours) in our experiments.

After obtaining the performance model coefficients, we proceed to run our \emph{iGniter} strategy in Alg.~\ref{alg-inference-provisioning} on a p3.2xlarge EC2 instance. The computation overhead and memory consumption of \emph{iGniter} are negligible, which are merely $3.64$ milliseconds and $53.17$ MB, respectively. As the number of workloads is increased to $1,000$ shown in Fig.~\ref{fig-evaluation-scalability}, the computation overhead is still within $4.61$ seconds and the memory overhead is less than $55$ MB. This is because the computation time and memory consumption of Alg.~\ref{alg-inference-provisioning} are quadratic to and linear to the number of DNN inference workloads, respectively, as analyzed in Sec.~\ref{sec:design-algorithm}. As a result, the runtime overhead of our \emph{iGniter} strategy can be acceptable in practice.

\section{Related Work}
\label{sec:related}

\textbf{Achieving Predictable DNN Inference on GPUs.} As summarized in Table~\ref{table-comparsion}, there have been a number of works on guaranteeing DNN inference performance SLOs on GPUs. In the scenario of \textbf{\emph{disabling GPU sharing}} (\emph{i.e.,} a GPU serves one DNN inference at a time), Clipper~\cite{DXGMJI2017} proposes caching, adaptive batch size, and dynamic model selection techniques to achieve low-latency and high-throughput DNN inference. BatchDVFS~\cite{SSM2022} combines adaptive batching with the DVFS technique to maximize the inference request throughput while guaranteeing the power caps.

\begin{table}[!t]\vspace{+6pt}
\renewcommand{\arraystretch}{1.5}
\centering
\caption{Comparison of predictable DNN inference systems on GPUs.}
\label{table-comparsion}
\resizebox{\linewidth}{!}{
\begin{tabular}{c|ccccc}
\toprule[1pt]
  \multirow{2}{*}{\textbf{Strategies}} & Interference & Spatial & Profiling & Workload & \multirow{2}{*}{Batching}\\
  & awareness & sharing & overhead & placement\\
\midrule[1pt]
Clipper~\cite{DXGMJI2017}  & \XSolid & \XSolid & N/A & \XSolid & \Checkmark\\
\hline
BatchDVFS~\cite{SSM2022} & \XSolid & \XSolid & lightweight & \XSolid & \Checkmark\\
\hline
Nexus~\cite{HLYLBMAR2019} & \XSolid & \XSolid & lightweight & \Checkmark & \Checkmark\\
\hline
Clockwork~\cite{ARSWAYJ2020} & \XSolid & \XSolid & lightweight & \Checkmark & \Checkmark\\
\hline
Morphling~\cite{LLYWBXJL2021} & \XSolid & \XSolid & lightweight & \XSolid & \Checkmark\\
\hline
Cocktail~\cite{JCPMC2022} & \XSolid & \XSolid & lightweight & \XSolid & \XSolid\\
\hline
INFaaS~\cite{FQNC2021} & \Checkmark & \XSolid & lightweight & \Checkmark & \Checkmark\\
\hline
Scrooge~\cite{YRR2021} & \XSolid & multiple & heavy & \Checkmark & \Checkmark\\
\hline
MIG-serving~\cite{CZJYSZYC2021} & \XSolid & multiple & heavy & \Checkmark & \Checkmark\\
\hline
INFless~\cite{YLYHJMXK2022} & \XSolid & multiple & lightweight & \Checkmark & \Checkmark\\
\hline
GSLICE~\cite{ASK2020} & \XSolid & multiple & N/A & \XSolid & \Checkmark\\
\hline
gpu-lets~\cite{SSYJYJ2022} & \Checkmark & 2 & heavy & \Checkmark & \Checkmark\\
\hline
\emph{\textbf{iGniter}} & \Checkmark & multiple & lightweight & \Checkmark & \Checkmark\\
\bottomrule[1pt]
\end{tabular}\vspace{-10pt}}
\end{table}

In the scenario of \textbf{\emph{temporal sharing}} of GPUs, Nexus~\cite{HLYLBMAR2019} proposes batching-aware scheduling based on Clipper~\cite{DXGMJI2017} to improve the GPU utilization. Clockwork~\cite{ARSWAYJ2020} designs fine-grained request-level scheduling to order user requests based on their latency SLOs. Morphling~\cite{LLYWBXJL2021} utilizes meta-learning to quickly configure the batch size, CPU cores, GPU memory, GPU timeshare, and GPU type for each inference workload. While sharing the adaptive batching and workload placement techniques with the prior works above, \emph{iGniter} aims to \emph{cost-efficiently} guarantee the performance SLOs based on GPU \emph{spatial sharing}, instead of maximizing the request throughput of inference workloads. To further reduce the monetary cost of DNN inference, two more recent works (\emph{i.e.,} Cocktail~\cite{JCPMC2022}, INFaaS~\cite{FQNC2021}) design the heterogeneous instance/accelerator selection, resource autoscaling, and dynamic model-variants selection techniques for cost-effective resource provisioning. These techniques above can be incorporated into \emph{iGniter} to further save the inference budget. In addition, our \emph{SM-level} resource scaling in \emph{iGniter} (\emph{i.e.,} $r_{unit}$ in Algorithm~\ref{alg-resource-allocation}) is more \emph{fine-grained} than the \emph{device-level} resource scaling in Cocktail and INFaaS.

In the scenario of \textbf{\emph{spatial sharing}} of GPUs, Scrooge~\cite{YRR2021} leverages the CUDA streams and batching techniques to pack DNN inference on VMs to ensure the performance SLOs of media applications. Using the latest multi-instance GPU (MIG)~\cite{mig} featured A100 GPUs, MIG-serving~\cite{CZJYSZYC2021} optimizes a set of GPU partitions and DNN inference deployments to meet performance SLOs. To further maximize the request throughput, INFless~\cite{YLYHJMXK2022} adopts batching and heterogeneous CPU-GPU resources for DNN inference in the serverless platform. GSLICE~\cite{ASK2020} and gpu-lets~\cite{SSYJYJ2022} \emph{separately} adjust the batch size and allocated GPU resources for inference workloads. However, the prior works above are mostly oblivious to performance interference and thus they tend to cause long-tail latency due to the severe GPU resource contention. In contrast. \emph{iGniter} proactively considers (\emph{i.e.,} minimizes) the performance interference among co-located inference workloads and \emph{jointly} optimizes the GPU resource allocation and batch size configuration. 

\textbf{Modeling Performance Interference in Clouds.} There have been prior works on modeling the performance interference~\cite{FFLHBB2014} and hardware heterogeneity~\cite{FFH2016} of cloud CPU instances. For instance, VELTAIR~\cite{ZJZQCM2022} builds a simple linear interference model using L3 cache miss rate and L3 access statistics. To particularly model the performance interference among co-located VMs based on \textbf{\emph{temporal sharing}} of GPUs, Xu \emph{et al.}~\cite{XNMMR2019} build a random forest regression model with a set of factors such as GPU/memory utilization and the average kernel length. As DNN training and inference workloads become prevailing in the cloud~\cite{HFLZF2019}, Horus~\cite{GDRARP2021} leverages GPU utilization to estimate the performance interference among co-located \emph{DNN training} jobs through fitting a quadratic function, while \emph{iGniter} focuses on modeling the \emph{DNN inference} performance using a set of easily-accessible GPU system and workload metrics.

Different from the interference above caused by the \emph{context switching of temporal sharing} of GPUs, NVIDIA MPS allows DNN inference to \textbf{\emph{spatially share}} GPU resources. To model the interference caused by GPU resource contention, Prophet~\cite{QHMRJL2017} characterizes the contention of GPU processing elements and DRAM bandwidth~\cite{WQKNZJM2022} as well as PCIe bandwidth in the \emph{default} mode of MPS~\cite{QHJL2016}. Based on the \emph{MPS with limited GPU resources}, gpu-lets~\cite{SSYJYJ2022} builds a linear regression model using the L2 cache and DRAM bandwidth utilization to predict the latency increases for only \emph{two} inference workloads. However, it requires profiling a number (\emph{e.g.,} thousands) of possible workload configurations, which brings heavy runtime overhead. Different from the models above, \emph{iGniter} builds an analytical model to predict the interference among multiple (\emph{i.e.,} more than $2$) inference workloads by a \emph{lightweight} workload profiling with a limited number (\emph{i.e.,} $11$) of configurations. Moreover, our \emph{iGniter} model \emph{comprehensively} considers the severe contention of GPU scheduler, L2 GPU cache space, and GPU power consumption among co-located inference workloads.

\section{Conclusion and Future Work}
\label{sec:conclusion}

This paper presents the design and implementation of \emph{iGniter}, an interference-aware GPU resource provisioning framework for achieving predictable DNN inference in the cloud. By leveraging the key system and workload metrics, we first devise a lightweight analytical performance model to capture the performance interference of inference workloads co-located on GPUs. Such a performance model further guides the design of a cost-efficient GPU resource provisioning strategy in \emph{iGniter}. It jointly optimizes the GPU resource allocation and batch size configuration to greedily minimize the performance interference of DNN inference workloads. Extensive prototype experiments on Amazon EC2 demonstrate that \emph{iGniter} can guarantee the performance SLOs of cloud-based DNN inference workloads, while saving the monetary cost by up to $25\%$ compared with the state-of-the-art resource provisioning strategies.

We plan to extend \emph{iGniter} in the following directions: (1) provisioning  DNN inference workloads with multiple types of GPU hardware or accelerators, (2) allocating multiple GPU instances to a DNN inference workload with an extremely large request arrival rate, (3) negotiating the tradeoff between minimizing the monetary cost and maximizing the performance of DNN inference workloads, (4) deploying a dynamic temporal and spatial GPU sharing strategy for time-varying request arrival rates, and (5) examining the effectiveness of \emph{iGniter} in the mixed deployment scenario of DNN inference and training workloads.

% or


%\ifCLASSOPTIONcompsoc
%\else
%  % regular IEEE prefers the singular form
%\fi

%The corresponding author is Fei Xu (fxu@cs.ecnu.edu.cn). This work was supported in part by the NSFC under grant No.61972158, in part by the Science and Technology Commission of Shanghai Municipality under grant No.20511102802 and No.18DZ2270800, in part by the Natural Science Foundation of Shanghai under grant NO.21ZR1419900, and in part by the Tencent Corporation.

\ifCLASSOPTIONcaptionsoff
  \newpage
\fi

\bibliographystyle{IEEEtran}
\bibliography{ref}

\vspace{-60pt}
\begin{IEEEbiography}[{\includegraphics[width=1in,height=1.25in,clip,keepaspectratio]{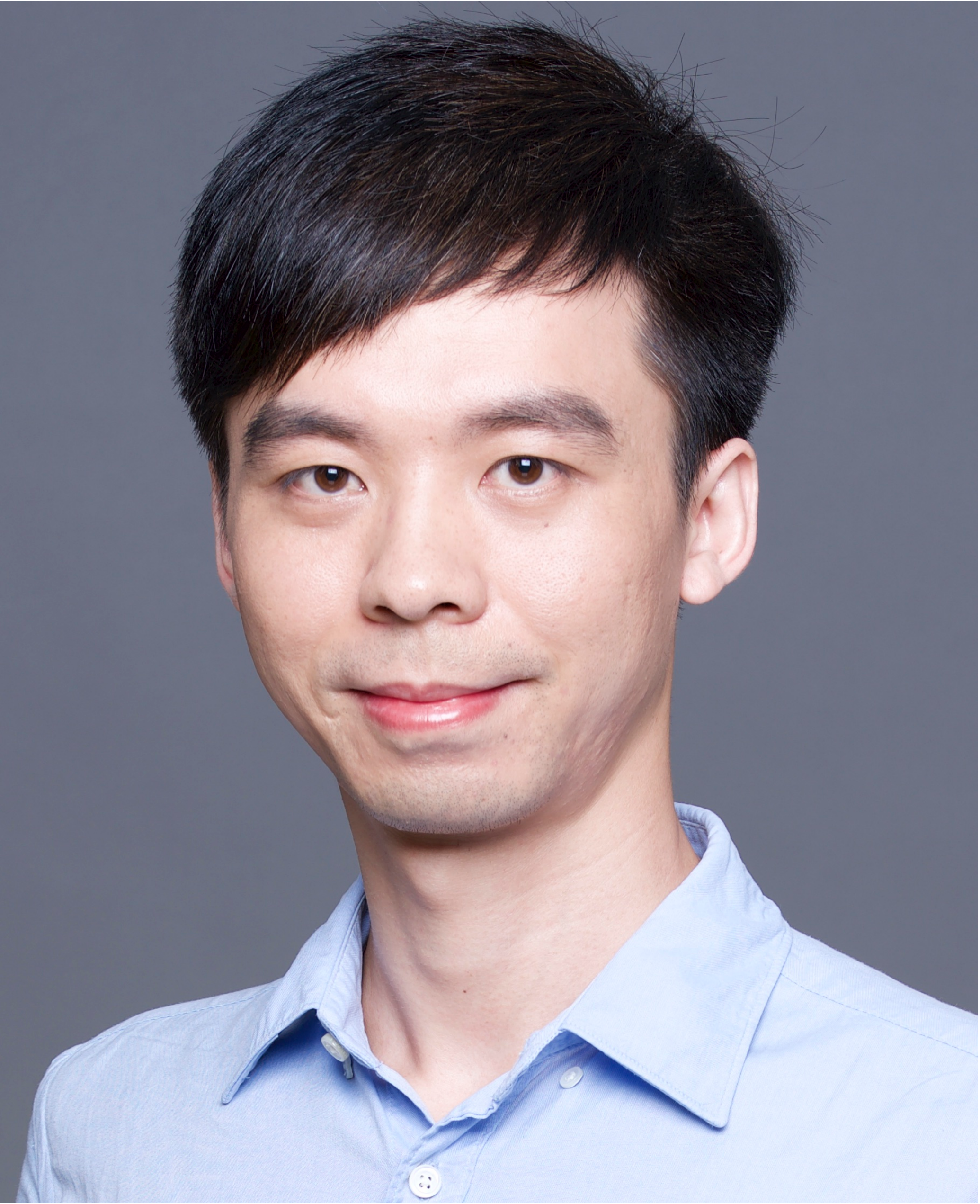}}]{Fei Xu}
received the B.S., M.E., and Ph.D. degrees in 2007, 2009, and 2014, respectively, all from the Huazhong University of Science and Technology (HUST), Wuhan, China. He received Outstanding Doctoral Dissertation Award in Hubei province, China, and ACM Wuhan \& Hubei Computer Society Doctoral Dissertation Award in 2015. He is currently an associate professor with the School of Computer Science and Technology, East China Normal University, Shanghai, China. His research interests include cloud computing and datacenter, virtualization technology, and distributed systems.
\end{IEEEbiography}

\vspace{-50pt}
\begin{IEEEbiography}[{\includegraphics[width=1in,height=1.25in,clip,keepaspectratio]{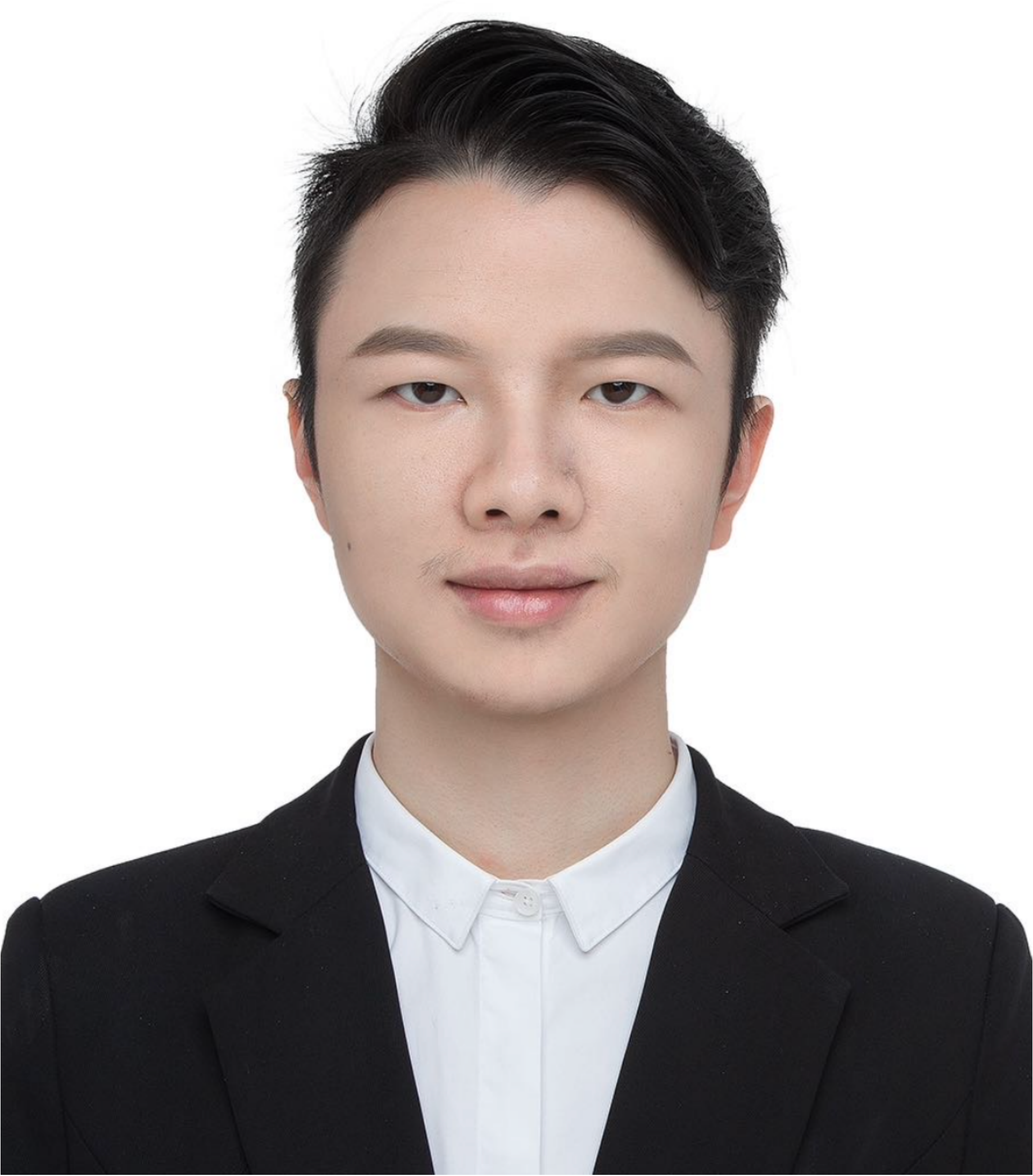}}]{Jianian Xu}
received his B.S. degree in Polymer Materials and Engineering from Qingdao University of Science and Technology in 2019. He is currently working toward the master's degree in the School of Computer Science and Technology, East China Normal University, Shanghai, China. His research interests focus on cloud computing and distributed machine learning systems.
\end{IEEEbiography}

\vspace{-50pt}
\begin{IEEEbiography}[{\includegraphics[width=1in,height=1.25in,clip,keepaspectratio]{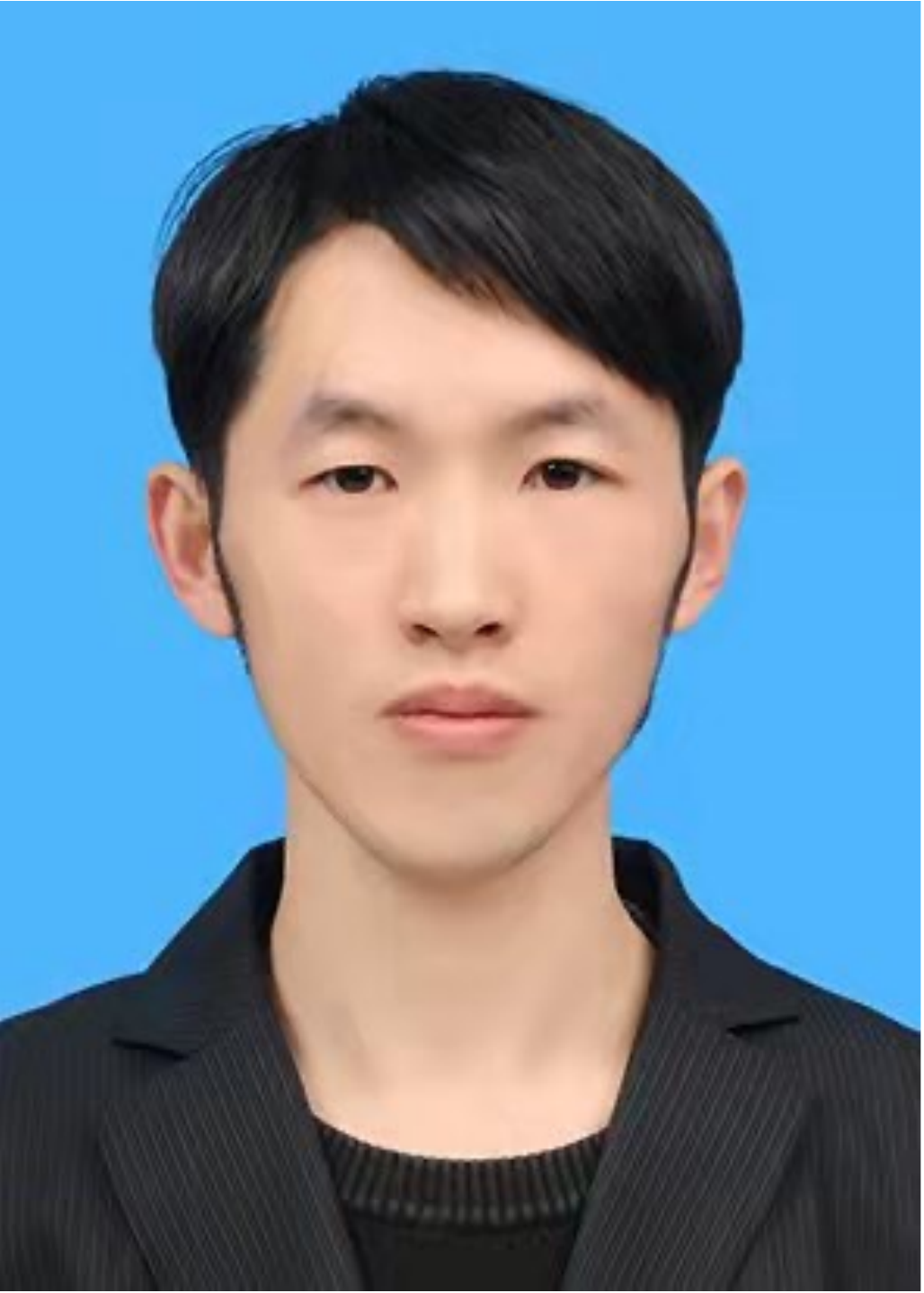}}]{Jiabin Chen}
received his B.S. degree in Optoelectronic Information Science and Engineering from Harbin Institute of Technology, Weihai in 2019. He is currently working toward the master's degree in the School of Computer Science and Technology, East China Normal University, Shanghai, China. His research interests focus on cloud computing and distributed machine learning systems.
\end{IEEEbiography}

\vspace{-50pt}
\begin{IEEEbiography}[{\includegraphics[width=1in,height=1.25in,clip,keepaspectratio]{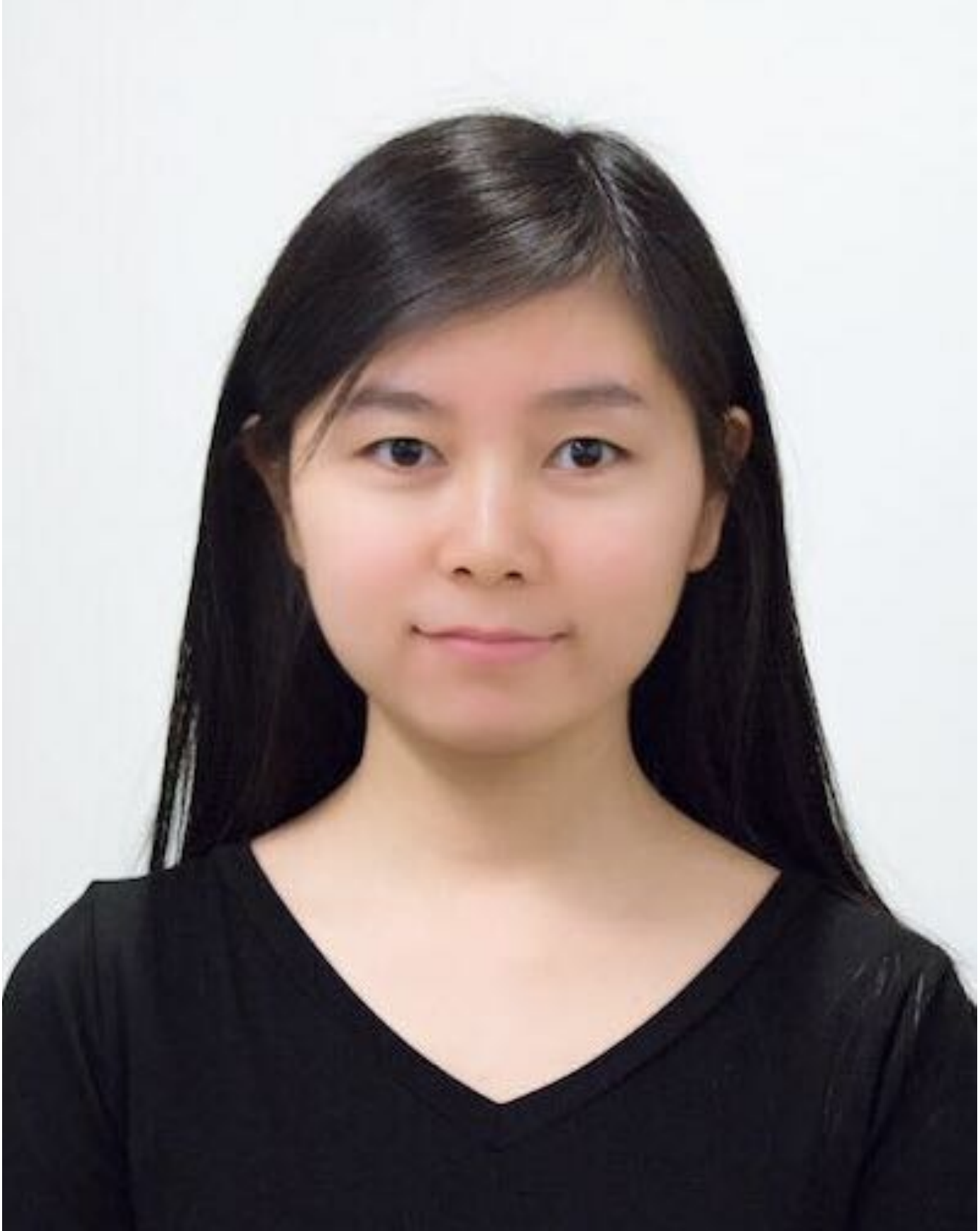}}]{Li Chen}
received the BEngr degree from the Department of Computer Science and Technology, Huazhong University of Science and Technology, China, in 2012 and the MASc degree from the Department of Electrical and Computer Engineering, University of Toronto, in 2014 and the PhD degree in computer science and engineering from the Department of Electrical and Computer Engineering, University of Toronto, in 2018. She is currently an assistant professor with the Department of Computer Science, School of Computing and Informatics, University of Louisiana at Lafayette, Lafayette, USA. Her research interests include big data analytics systems, cloud computing, datacenter networking, and resource allocation.
\end{IEEEbiography}

\vspace{-50pt}
\begin{IEEEbiography}[{\includegraphics[width=1in,height=1.25in,clip,keepaspectratio]{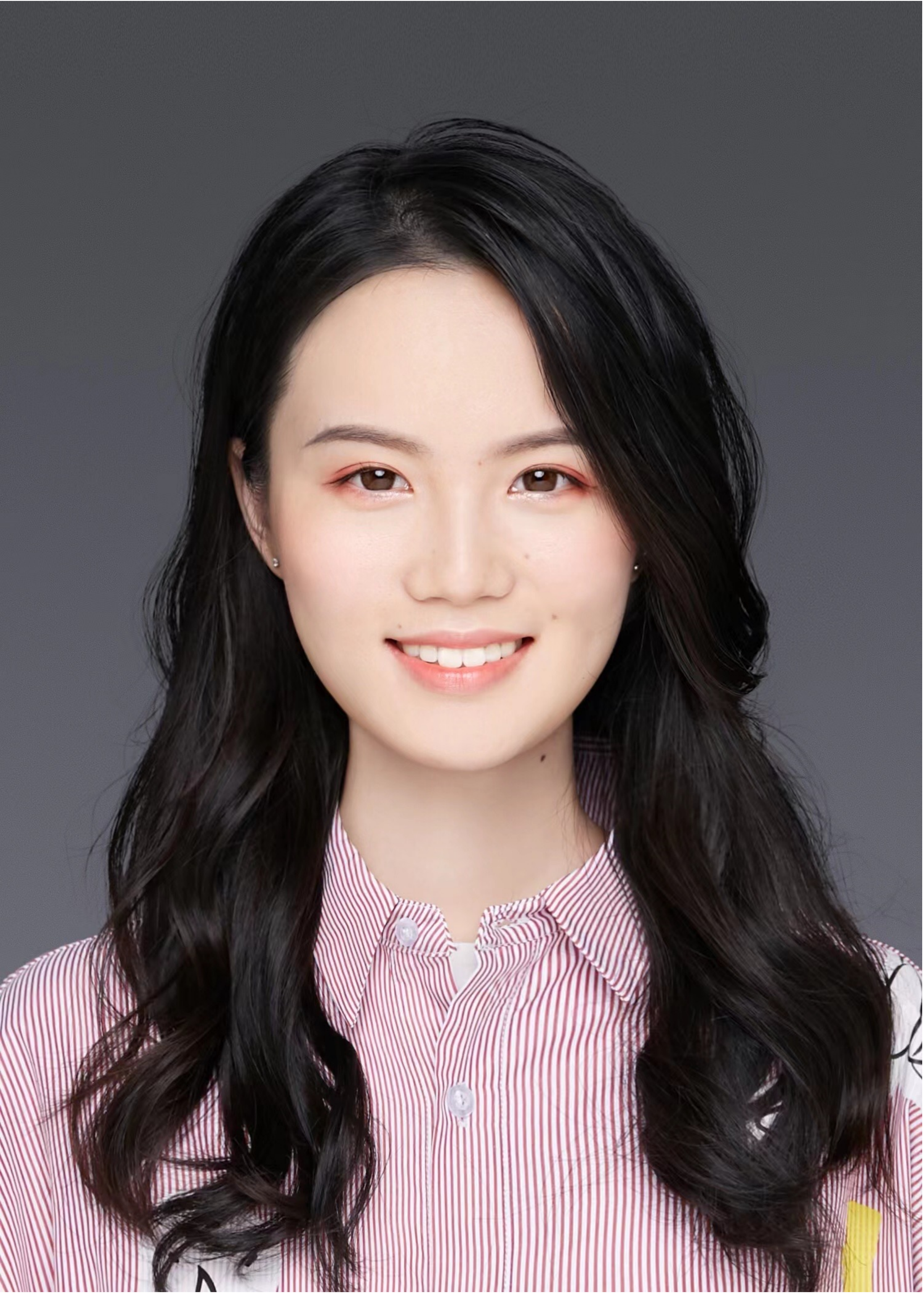}}]{Ruitao Shang}
received her B.S. degree in Computer Science from East China Normal University (ECNU) in 2020. She is currently pursuing her MS degree in Computer Science in the School of Computer Science and Technology at ECNU. Her current research interests focus on cloud computing and distributed machine learning systems.
\end{IEEEbiography}

\vspace{-50pt}
\begin{IEEEbiography}[{\includegraphics[width=1in,height=1.25in,clip,keepaspectratio]{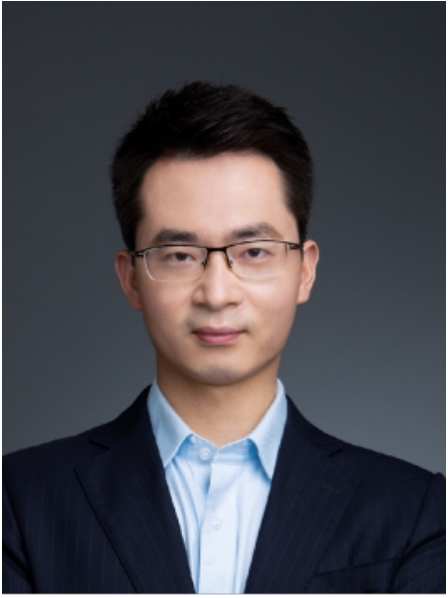}}]{Zhi Zhou}
received the B.S., M.E., and Ph.D. degrees in 2012, 2014, and 2017, respectively, all from the School of Computer Science and Technology at Huazhong University of Science and Technology (HUST), Wuhan, China. He is currently an associate professor in the School of Computer Science and Engineering at Sun Yat-sen University, Guangzhou, China. In 2016, he was a visiting scholar at University of G{\"o}ttingen. He was nominated for the 2019 CCF Outstanding Doctoral Dissertation Award, the sole recipient of the 2018 ACM Wuhan \& Hubei Computer Society Doctoral Dissertation Award, and a recipient of the Best Paper Award of IEEE UIC 2018. His research interests include edge computing, cloud computing, and distributed systems.
\end{IEEEbiography}

\vspace{-50pt}
\begin{IEEEbiography}[{\includegraphics[width=1in,height=1.25in,clip,keepaspectratio]{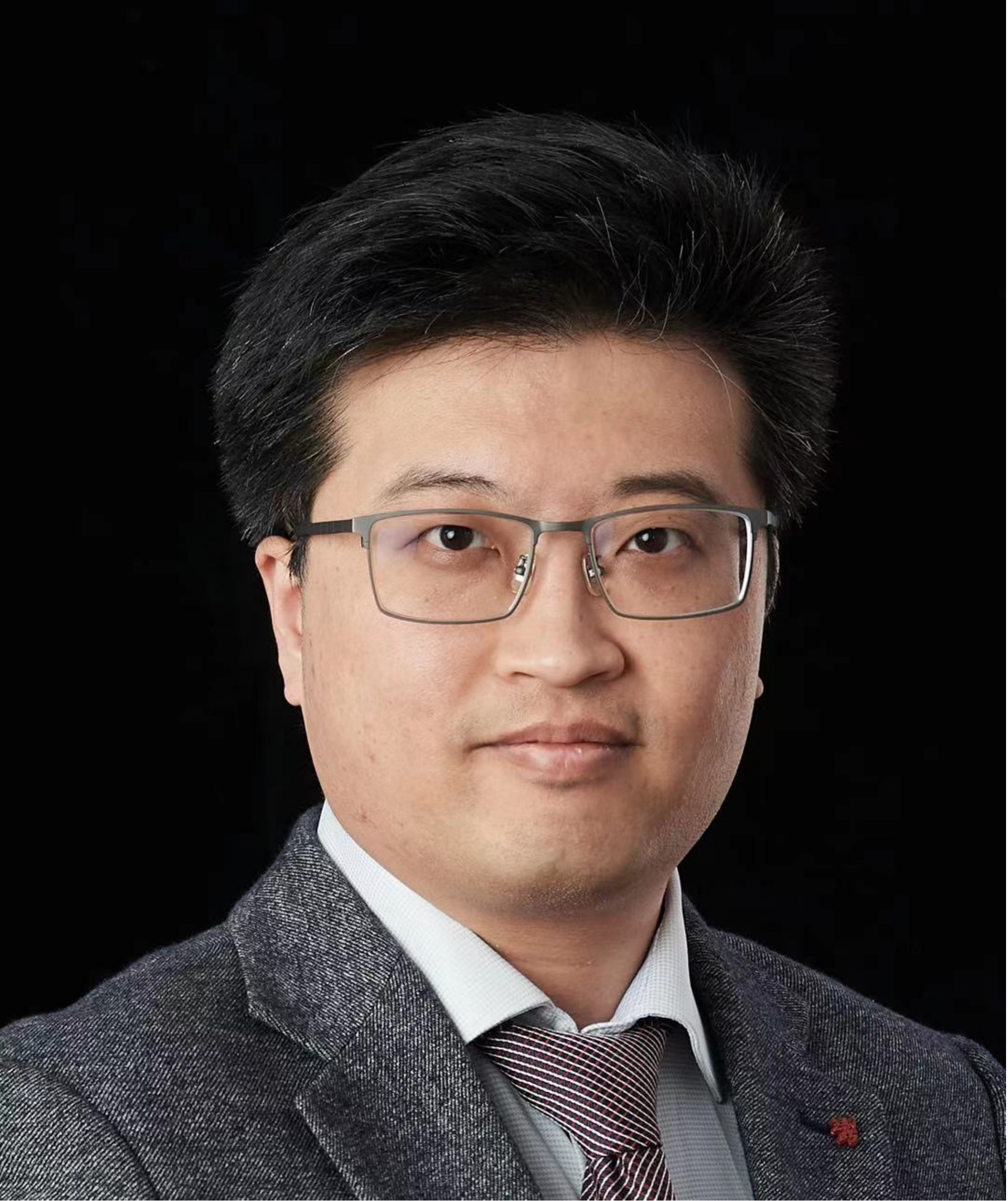}}]{Fangming Liu}
(S'08, M'11, SM'16) received the B.Eng. degree from the Tsinghua University, Beijing, and the Ph.D. degree from the Hong Kong University of Science and Technology, Hong Kong. He is currently a Full Professor with the Huazhong University of Science and Technology, Wuhan, China. His research interests include cloud computing and edge computing, datacenter and green computing, SDN/NFV/5G and applied ML/AI. He received the National Natural Science Fund (NSFC) for Excellent Young Scholars, and the National Program Special Support for Top-Notch Young Professionals. He is a recipient of the Best Paper Award of IEEE/ACM IWQoS 2019, ACM e-Energy 2018 and IEEE GLOBECOM 2011, the First Class Prize of Natural Science of Ministry of Education in China, as well as the Second Class Prize of National Natural Science Award in China.
\end{IEEEbiography}

\newpage
\clearpage
\appendix

\section{Proof of Theorem~\ref{thm-lower-upper-bound}}
\label{sec:appendix}

\begin{proof}
We first calculate the appropriate batch size $b_{appr}^{i}$ that just meets the arrival rate. Specifically, we substitute Eq.~(\ref{eq-inference-latency}) into Constraint~(\ref{eq-cons-slo}), yielding $\frac{T_{slo}^{i}}{2} - t_{load}^{i} - t_{feedback}^{i} \geq t_{gpu}^{ij}$ when an inference workload $i$ is running on a GPU $j$. Given a batch size, the GPU execution latency increases as the amount of allocated GPU resources decreases. Accordingly, in order to minimize the amount of GPU resources, we set the GPU execution latency to the maximum value as
\begin{equation}\label{eq-transformed-constraint}
	t_{gpu}^{ij} = \frac{T_{slo}^{i}}{2} - t_{load}^{i} - t_{feedback}^{i}.
\end{equation}
By substituting Eq.~(\ref{eq-transformed-constraint}), Eq.~(\ref{eq-throughput}), and Eq.~(\ref{eq-pcie-latency}) into Constraint~(\ref{eq-cons-arrivalrate}), we have $b^{i} \geq  \frac{T_{slo}^{i} \cdot R^{i} \cdot B_{pcie}}{2 \cdot (B_{pcie} + \cdot R^{i} \cdot d_{load}^{i})} $. In addition, a larger batch size generally indicates a higher GPU execution latency given an amount of allocated GPU resources. Accordingly, we simply choose the \emph{appropriate} batch size $b_{appr}^{i}$ that \emph{just meets} the arrival rate, which is given by
\begin{equation}
    b_{appr}^{i} = \bigg\lceil \frac{T_{slo}^{i} \cdot R^{i} \cdot B_{pcie}}{2 \cdot (B_{pcie} + R^{i} \cdot d_{load}^{i})} \bigg\rceil.\nonumber
\end{equation}
In more detail, if we increase the batch size $b_{appr}^{i}$, the GPU resources allocated to the workload $i$ requires increasing. Otherwise, if we reduce the batch size $b_{appr}^{i}$, it will violate Constraint~(\ref{eq-cons-arrivalrate}) (\emph{i.e.,} the request arrival rate cannot be guaranteed). Accordingly, we consider $b_{appr}^{i}$ as the \emph{appropriate} batch size for our optimization problem.

We next obtain the lower bound $r_{lower}^{i}$ of GPU execution resources for each workload $i$ as follows. By substituting $b_{appr}^{i}$, Eq.~(\ref{eq-inference-latency}), Eq.~(\ref{eq-pcie-latency}), Eq.~(\ref{eq-gpu-latency}),  Eq.~(\ref{eq-schedule-latency}), Eq.~(\ref{eq-increased-schedule-latency}), Eq.~(\ref{eq-cache-interference}), and Eq.~(\ref{eq-single-gpu-latency}) into Constraint~(\ref{eq-cons-slo}), we calculate the amount of allocated resources $r^{ij}$ on a GPU device as below,
\begin{equation}
	\begin{split}
    r^{ij} & \geq \frac{k_{1}^{i} \cdot (b_{appr}^{i})^2 + k_{2}^{i} \cdot b_{appr}^{i} + k_{3}^{i}}{\Big(\frac{T_{slo}^{i}}{2} - \frac{(d_{load}^{i} + d_{feedback}^{i}) \cdot b_{appr}^{i}}{B_{pcie}}\Big) \cdot \frac{f^{j}}{F} - k_{5}^{i} - k_{sch}^{i} \cdot n_{k}^{i}} - k_{4}^{i} \\
    & \geq \frac{k_{1}^{i} \cdot (b_{appr}^{i})^2 + k_{2}^{i} \cdot b_{appr}^{i} +k_{3}^{i}}{\frac{T_{slo}^{i}}{2} -\frac{(d_{load}^{i} + d_{feedback}^{i}) \cdot b_{appr}^{i}}{B_{pcie}} -k_{5}^{i}-k_{sch}^{i} \cdot  n_{k}^{i}}-k_{4}^{i}. \nonumber
    \end{split}
\end{equation}
As the GPU resources are allocated in units of $r_{unit}$ which is set as $2.5\%$ for NVIDIA V100 GPUs, the lower bound $r_{lower}^{i}$ of GPU execution resources for each workload $i$ can be calculated by
\begin{equation}
    r_{lower}^{i} = \bigg\lceil \frac{\gamma^{i}}{\delta^{i} \cdot r_{unit}} - \frac{k_{4}^{i}}{r_{unit}} \bigg\rceil \cdot r_{unit}, \nonumber
\end{equation}
where $\gamma^{i}=k_{1}^{i} \cdot (b_{appr}^{i})^2 + k_{2}^{i} \cdot b_{appr}^{i} +k_{3}^{i}$ and $\delta^{i} = \frac{T_{slo}^{i}}{2} -\frac{(d_{load}^{i} + d_{feedback}^{i}) \cdot b_{appr}^{i}}{B_{pcie}}-k_{5}^{i}-k_{sch}^{i}\cdot n_{k}^{i}$.
\end{proof}

\end{document}